\documentclass[lettersize,journal]{IEEEtran}
\usepackage{cite}
\usepackage{bm}
\usepackage{amsmath,amsfonts}
\usepackage[noend]{algorithmic}
\usepackage{algorithm}
\usepackage{array}
\usepackage{subfig}
\usepackage{textcomp}
\usepackage{stfloats}
\usepackage{url}
\usepackage{verbatim}
\usepackage{graphicx}
\usepackage[font=small]{caption}  
\usepackage{booktabs} 
\usepackage{makecell}
\usepackage{multirow}
\usepackage{stfloats}
\usepackage{hyperref}

\usepackage{threeparttable}

\begin{document}


\title{NOC4SC: A Bandwidth-Efficient Multi-User Semantic Communication Framework for Interference-Resilient Transmission}

\author{
Yunhao Wang,~\IEEEmembership{Graduate Student Member,~IEEE,}
Shuai Ma,~\IEEEmembership{Member,~IEEE,}
Pengfei He,~\IEEEmembership{Graduate Student Member,~IEEE,}
Dahua Gao,~\IEEEmembership{Member,~IEEE,}
Guangming Shi,~\IEEEmembership{Fellow,~IEEE,}
and Xiang Cheng,~\IEEEmembership{Fellow,~IEEE}

\thanks{A preliminary version of this work has been accepted for presentation at the 2025 IEEE Global Communications Conference (GLOBECOM), Taipei, Taiwan, December 8, 2025.}
\thanks{This work was supported in part by the National Science and Technology Major Project - Mobile Information Networks under Grant No.2024ZD1300700, and in part by the Natural Science Foundation (NSF) of China No.62293483 and No.62476206.}
\thanks{Yunhao Wang is with the School of Electronic and Computer Engineering, Peking University, Shenzhen 518055, China, and also with the Department of Networked Intelligence, Peng Cheng Laboratory, Shenzhen 518066, China (e-mail: yunhaowang@stu.pku.edu.cn)}
\thanks{Shuai Ma is with the Department of Networked Intelligence, Peng Cheng Laboratory, Shenzhen 518066, China. (e-mail: mash01@pcl.ac.cn).}
\thanks{Pengfei He is with the School of Artificial Intelligence, Xidian University, Xi’an, Shaanxi 710071, China, and also with the Department of Networked Intelligence, Peng Cheng Laboratory, Shenzhen 518066, China (e-mail: hepengfei@stu.xidian.edu.cn)}
\thanks{Dahua  Gao is with the Key Lab of Intelligent Perception and Image Understanding of Ministry of Education, Xidian University, Xi’an, 710071, China (e-mail: dhgao@xidian.edu.cn).}
\thanks{Guangming Shi \textit{(Corresponding Author)} is with the School of Artificial Intelligence, Xidian University, Xi’an, Shaanxi 710071, China, and also with the Department of Networked Intelligence, Peng Cheng Laboratory, Shenzhen, 518066, China (e-mail: shigm@pcl.ac.cn).}
\thanks{Xiang Cheng is with the State Key Laboratory of Photonics and Communications, School of Electronics, Peking University, Beijing 100871, China (e-mail: xiangcheng@pku.edu.cn).}
\thanks{The code will be released on \url{https://github.com/WYHxuebi/NOC4SC}.}
}



\maketitle

\begin{abstract}
With the explosive growth of connected devices and emerging applications, current wireless networks are encountering unprecedented demands for massive user access, where the inter-user interference has become a critical challenge to maintaining high quality of service (QoS) in multi-user communication systems. To tackle this issue, we propose a bandwidth-efficient semantic communication paradigm termed Non-Orthogonal Codewords for Semantic Communication (NOC4SC), which enables simultaneous same-frequency transmission without spectrum spreading. By leveraging the Swin Transformer, the proposed NOC4SC framework enables each user to independently extract semantic features through a unified encoder–decoder architecture with shared network parameters across all users, which ensures that the user's data remains protected from unauthorized decoding. Furthermore, we introduce an adaptive NOC and SNR Modulation (NSM) block, which employs deep learning to dynamically regulate SNR and generate approximately orthogonal semantic features within distinct feature subspaces, thereby effectively mitigating inter-user interference. Extensive experiments demonstrate the proposed NOC4SC achieves comparable performance to the DeepJSCC-PNOMA and outperforms other multi-user SemCom baseline methods.
\end{abstract}


\section{Introduction}
\IEEEPARstart{W}{ith} the rapid development of next-generation wireless technologies, traditional communication systems are struggling to meet the growing demands for massive connectivity \cite{ref1}. Emerging applications-such as massive IoT \cite{ref2}, autonomous driving \cite{ref3}, digital twins \cite{ref4}, remote healthcare \cite{ref5}, and smart cities \cite{ref6}-necessitate the support of an unprecedented number of simultaneous device connections. This explosive growth in connection density exacerbates inter-user interference and places significant pressure on limited spectrum resources \cite{ref7}. 

Unlike conventional systems that emphasize bit-level accuracy, semantic communication (SemCom) represents a paradigm shift from transmitting exact symbols to conveying meaningful information for the receiver’s understanding or task execution \cite{ref10}. In this context, semantics refers not to pixel-wise precision but to the preservation of task-relevant meaning embedded in the transmitted content, which enables the receiver to correctly infer or reconstruct the intended message at the semantic level, even under symbol-level distortion. Therefore, SemCom significantly reduces bandwidth consumption and enhances reliability in noisy or resource-limited environments \cite{ref12, ref14}. These advantages are particularly valuable in multi-user scenarios, where limited spectral resources and dense connectivity exacerbate interference. Consequently, developing efficient SemCom that support robust multi-user access has emerged as an effective direction to address these challenges in the wireless networks.

\captionsetup[table]{justification=centering, labelsep=space, textfont=sc} 
\begin{table*}[t]
    \renewcommand\arraystretch{2.0}
    \setlength{\tabcolsep}{2.5pt}
    \caption{ \\ Contrasting our Contributions to the Literature \label{tab9}}
    \centering
    \begin{tabular}{c|c|c|c|c|c|c|c}
        \hline
        \hline
        \textbf{Technique} & \textbf{Method} & \textbf{Access} & \textbf{Coding} & \textbf{Modality} & \makecell{\textbf{Bandwidth} \\ \textbf{Efficiency}}  & \makecell{\textbf{Param. and Arch.} \\ \textbf{Sharing}} & \textbf{Scenery} \\
        \hline
        \multirow{2}{*}{Model-driven} 
         &  CDMA \cite{ref31} & Code-domain & Separation-based & Bits & $\bm{\times}$ & \textbf{-} & Uplink / Downlink \\
         & NOMA \cite{ref33} & Power-domain & SC + SIC & Bits & $\bm{\surd}$ & \textbf{-} & Uplink / Downlink \\
        \hline
        \multirow{8}{*}{Learning-based} 
         & DeepSIC \cite{shlezinger2020deepsic} & Power-domain & Separation-based & Bits & $\bm{\surd}$ & $\bm{\times}$ & Uplink \\
         & NOMASC \cite{li2023non} & Power-domain & JSCC & Images & $\bm{\surd}$ & $\bm{\times}$ & Downlink \\
         & DeepSC-MT \cite{xie2022task} & Feature-domain & JSCC & Images, texts & $\bm{\surd}$ & $\bm{\times}$ & Uplink \\
         & DeepJSCC-NOMA \cite{yilmaz2023distributed} & Feature-domain & JSCC & Images & $\bm{\surd}$ & $\bm{\surd}$ & Uplink \\
         & DeepJSCC-PNOMA \cite{yilmaz2025learning} & Feature-domain & JSCC & Images & $\bm{\times}$ & $\bm{\surd}$ & Uplink \\
         & SFDMA \cite{shen2025semantic, ma2024semantic} & Feature-domain & JSCC & Images & $\bm{\surd}$ & $\bm{\times}$ & Uplink, Interference \\
         & \textbf{NOC4SC (Ours)} & \textbf{Feature-domain} & \textbf{JSCC} & \textbf{Images} & $\bm{\surd}$ & $\bm{\surd}$ & \textbf{Interference} \\
        \hline
        \hline
    \end{tabular}
    \begin{tablenotes}
        \footnotesize
        \item[*] \text{Noting:} \textbf{Arch.} denotes Architectures, and \textbf{Param.} denotes Parameters. \textbf{Bandwidth Efficiency} indicates whether spectrum spreading is applied. Since spreading increases bandwidth consumption, avoiding it leads to more efficient spectrum utilization. Since NOMA and CDMA are model-driven methods without neural network parameters, the entries under “Param. and Arch. Sharing” are marked as “–”.
    \end{tablenotes}
\end{table*}

\subsection{Related Work}

Efficient and reliable transmission in multi-user wireless networks is fundamentally limited by scarce spectrum resources and severe inter-user interference \cite{ref26, ref27, ref28}. The core technology of multiple access (MA) is the effective allocation of resources such as time, frequency, codewords and power resources. Most orthogonal multiple-access (OMA) schemes, as typical model-driven approaches—including time division multiple access (TDMA) \cite{ref29}, frequency division multiple access (FDMA) \cite{ref30}, and code division multiple access (CDMA) \cite{ref31}—achieve user separation by analytically allocating distinct orthogonal resources to each user, thereby minimizing inter-user interference. However, these meticulous allocation also leads to a paradoxical situation in which the limited orthogonal resources inherently restrict the number of users that can access the network simultaneously.

In contrast to OMA schemes, non-orthogonal multiple access (NOMA) enables multiple users to share the identical time-frequency resources. Most existing NOMA systems are primarily based on power-domain, where the transmitter assigns different power levels to users' signals based on their channel conditions\cite{ref33, shlezinger2020deepsic, li2023non}. At the receiver, successive interference cancellation (SIC) is employed, decoding user signals in descending order of their channel gains and sequentially removing the decoded signals \cite{ref33}. The superposition coding (SC) and SIC pairing forms the backbone of power-domain NOMA . Recent studies have also explored learning-based NOMA frameworks, which leverage neural networks (NN) to enhance power-domain multiplexing and inter-user interference mitigation. A data-driven MIMO detection framework, DeepSIC, was proposed in \cite{shlezinger2020deepsic}, which integrates deep NN into the iterative soft SIC to jointly learn symbol detection without requiring explicit CSI. In \cite{li2023non}, a NOMA-based multi-user SemCom system named NOMASC was proposed, integrating an asymmetric quantizer and NN to support heterogeneous users. However, power-domain NOMA still suffers from residual inter-user interference, especially when users have similar channel conditions—rendering power differentiation ineffective. Moreover, the sequential nature of SIC leads to increasing complexity and latency at the receiver.

Beyond power-domain approaches, recent learning-based multi-user SemCom frameworks have explored feature-domain representations to achieve semantic separation and mitigate semantic-level interference among users \cite{yilmaz2023distributed, yilmaz2025learning, xie2022task, shen2025semantic, ma2024semantic}. An uplink DeepJSCC-NOMA framework was proposed in \cite{yilmaz2023distributed}, employing parameter-sharing encoders and device embeddings to enable images transmission. However, this design depends on a fixed number of users, rendering the decoder architecture inflexible to network membership changes. The author in \cite{yilmaz2025learning} proposed an uplink DeepJSCC-PNOMA framework that introduces user-specific projection matrices to enable scalable image transmission. However, the projection-based design effectively performs a CDMA type spreading operation, where the spreading factor and the required bandwidth grow with the number of users, thereby reducing spectral efficiency. A task-oriented multi-user SemCom framework, DeepSC-MT, was proposed in \cite{xie2022task}, enabling end-to-end semantic transmission for multiple tasks, which preserves sentence-level meaning. In \cite{shen2025semantic, ma2024semantic}, the authors proposed SFDMA framework, where user features are encoded into distinguishable semantic subspaces to enable interference-resilient multi-user transmission. However, both frameworks depend on task- or user-specific network architectures, which hinder scalability across diverse users and tasks, while requiring individualized encoder–decoder parameters that increase model complexity and storage overhead.

\subsection{Motivation and Contribution}

Effective user separation remains a fundamental challenge due to the requirements of interference suppression, bandwidth efficiency, and scalability. Table \ref{tab9} offers a detailed comparison to the literature. Traditional model-driven code-domain schemes such as CDMA assign orthogonal spreading codes to ensure interference-free transmission. \textbf{\textit{However, the required hard-orthogonality-based spreading operations expand the occupied bandwidth in proportion to the user count, severely degrading spectral efficiency.}} We instead adopt a softer separation: predefined non-orthogonal codewords (NOCs) lightly modulate latent features so that separability emerges in the semantic space, without spectrum spreading. On the other hand, power-domain approaches such as NOMA \cite{ref33} and NOMASC \cite{li2023non} perform user separation through SC and SIC, \textbf{\textit{yet they suffer from residual inter-user interference and high decoding complexity}}.

\begin{figure*}[t]
\centering
\includegraphics[width=0.98\textwidth]{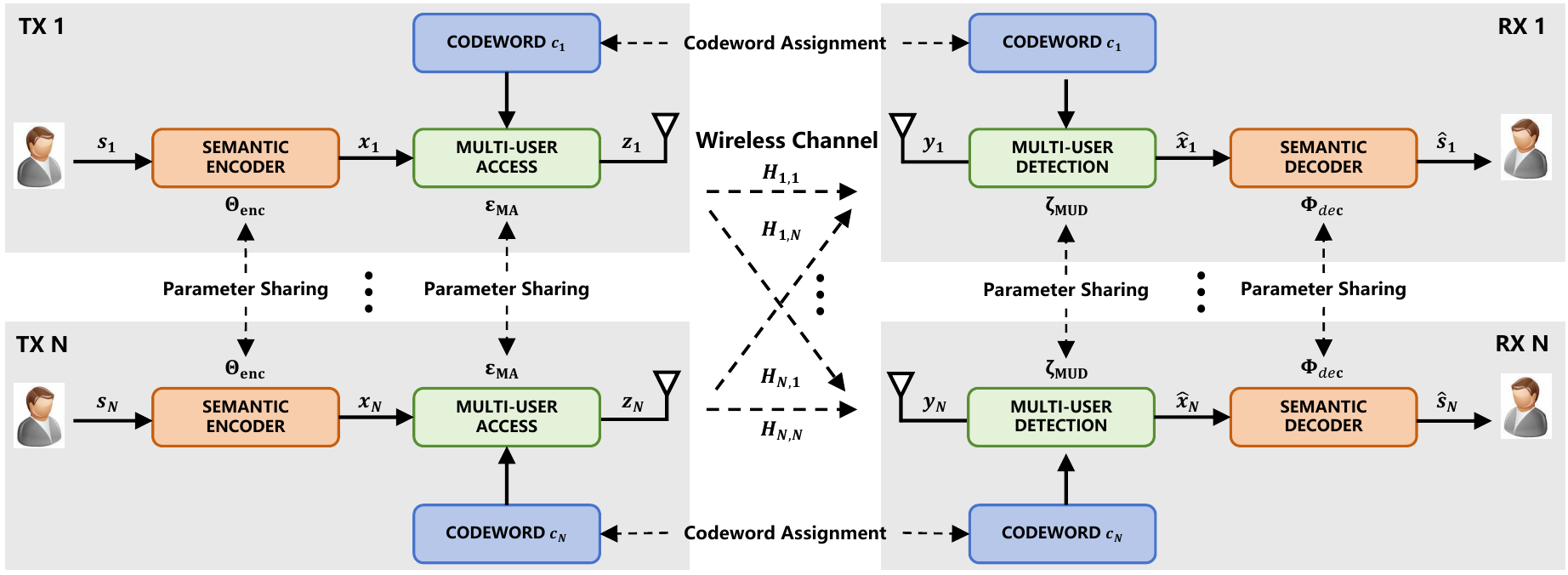}
\caption{The framework of multi-user interference system. The semantic encoder and decoder extract and reconstruct semantic information, while the multi-user access and detection modules modulate signals and mitigate inter-user interference.}
\label{fig1}
\end{figure*}

Learning-based representative works such as DeepJSCC-NOMA \cite{yilmaz2023distributed}, DeepJSCC-PNOMA \cite{yilmaz2025learning}, DeepSC-MT \cite{xie2022task}, and SFDMA \cite{shen2025semantic, ma2024semantic} achieve promising results in interference mitigation and semantic fidelity. \textbf{\textit{However, most existing frameworks depend on user or task specific architectures that require individualized encoder–decoder parameters, leading to substantial computational and storage overhead. Furthermore, projection-based method typically performs a CDMA-like spreading, causing bandwidth expansion and lower spectral efficiency as the number of users increases.}}

To this end, we propose NOC4SC—a Non-Orthogonal Codeword-based SemCom framework that realizes bandwidth-efficient and interference-resilient multi-user access under a unified encoder–decoder architecture. The proposed architecture adopts a symmetrical encoder–decoder design with shared parameters across users. The semantic encoder extracts high-level representations from images, which are subsequently modulated by the NOC and SNR Modulation (NSM) block. Each user is assigned a predefined NOC that serves as a unique semantic identifier, projecting its modulated features into distinguishable subspaces while maintaining approximate orthogonality across users-\textbf{\textit{without relying on spectrum spreading, power-domain separation, or task-dependent parameters}}. At the receiver, NOC4SC supports parallel multi-user decoding, substantially reducing complexity while enhancing resilience to inter-user interference. The main contributions can be summarized as follows:
\begin{itemize}
\item{A multi-user SemCom system based on NOCs named NOC4SC is proposed, which enables simultaneous semantic transmission over shared time-frequency resources. Each user independently reconstructs its semantic features through a unified encoder–decoder architecture with shared network parameters, effectively mitigating inter-user interference while ensuring user-level privacy. This unified design eliminates the need for user-specific models and facilitates flexible deployment across varying numbers of users.}
\item{A novel multi-user access mechanism is introduced, where each user is assigned a predefined NOC serving as its semantic identity. These NOCs softly project user features into distinguishable subspaces, achieving interference suppression without spectrum spreading, thereby improving bandwidth efficiency.}
\item{We design the NSM Block, which integrates user-specific NOC modulation with adaptive SNR control to generate approximately orthogonal semantic features without spread spectrum. Specifically, NOCs map semantic information into distinct subspaces for user separability, while the SNR adaptation mechanism improves robustness against dynamic noise conditions.}
\item{Extensive experimental results demonstrate that NOC4SC effectively achieves semantic feature separability and approximate orthogonality while reducing computational complexity, and consistently outperforms various multi-user semantic communication baselines.}
\end{itemize}

\subsection{Organization of This Article}

The organization of this article is as follows: Section II introduces the system model for the multi-user interference framework and the proposed NOC-based multi-user access mechanism. Section III presents the proposed NOC4SC system and numerical analysis are presented in Section IV. Finally, Section V concludes the paper.

\textit{Notation}: matrices, vectors and scalars are represented by bold uppercase letters, bold lowercase letters and lowercase letters, respectively. The sets of real and complex numbers are represented by \(\mathbb{R}\) and \(\mathbb{C}\), respectively. The notations used in this are summarized in Table \ref{tab10}.

\captionsetup[table]{justification=centering, labelsep=space, textfont=sc} 
\begin{table}[!h]
    \renewcommand\arraystretch{1.8}
    \setlength{\tabcolsep}{5pt}
    \caption{ \\ Table of abbreviations andnotations\label{tab10}}
    \centering
    \begin{tabular}{cl}
        \hline
        \hline
        \textbf{Notation} & \textbf{Description} \\
        \hline
        $\bm{s_i}$ & Input raw images of TX $i$  \\
        $\bm{x_i}$ & Semantic information vector of TX $i$ \\
        $\bm{z_i}$ & Semantic feature vector of TX $i$ \\
        $\bm{y_i}$ & Received signal over interference channel of RX $i$ \\
        $\bm{c_i}$ & Codeword vector assigned to the $i$-th user \\
        $f_{\text{enc}}(\, \bm{\cdot} \,;\, \bm{\Theta_{\textbf{enc}}} \,)$ & Semantic encoder in TX\\
        $g_{\text{NSM}}(\, \bm{\cdot} \,;\, \bm{\varepsilon_{\textbf{NSM}}} \,)$ & Multi-user access function of NSM Block \\
        $g_{\text{NSM}}(\, \bm{\cdot} \,;\, \bm{\zeta_{\textbf{NSM}}} \,)$ & Multi-user detection function of NSM Block \\
        $f_{\text{dec}}(\, \bm{\cdot} \,;\, \bm{\Phi_{\textbf{dec}}} \,)$ & Semantic decoder in RX \\
        \hline
        \hline
    \end{tabular}
\end{table}

\section{Multi-User Interference System Model with NOC-Based Access Mechanism}

In this section, we present a detailed description and modeling of the multi-user interference system. Without loss of generality, we consider the problem of image transmission in a multi-user SemCom system comprising \(N\) transmitters and \(N\) receivers, as illustrated in Fig.\ref{fig1}. 

\subsection{System Model}

The $i$-th user inputs the respective image $\bm{s_i} \in \mathbb{R}^{H \times W \times 3}$ to the semantic encoder, which transforms it into the semantic information $\bm{x_i} \in \mathbb{R}^{M_t \times 1}$. The encoder extracts salient features from the input image $\bm{s_i}$, reducing redundancy while preserving the essential semantic content required. This process can be formally expressed as:
\begin{equation}
    \bm{x_i} = f_{\text{enc}}(\bm{s_i} \, ; \, \bm{\Theta_{\textbf{enc}}}), \, i \in \{1, \ldots, N\},
    \label{eq1}
\end{equation}
where $f_{\text{enc}}(\, \bm{\cdot} \,;\, \bm{\Theta_{\textbf{enc}}} \,)$ represents the semantic encoding function, parameterized by $\bm{\Theta_{\textbf{enc}}}$. 

After semantic encoding, the semantic information $\bm{x_i}$ is mapped into the semantic feature $\bm{z_i} \in \mathbb{C}^{M_t \times 1}$ via the multi-user access mechanism. The feature $\bm{z_i}$ across different users are designed to be approximately orthogonal, ensuring distinguishable and separable. A detailed discussion of the NOC-based multi-user access mechanism is provided in Section III-A. The semantic feature can be represented as:
\begin{equation}
    \bm{z_i} = g_{\text{MA}}(\bm{x_i}, \, \bm{c_i} \,;\, \bm{\varepsilon_{\textbf{MA}}}),
    \label{eq2}
\end{equation}
where $g_{\text{MA}}(\, \bm{\cdot} \,;\, \bm{\varepsilon_{\textbf{MA}}} \,)$ represents the multi-user access mapping function, parameterized by $\bm{\varepsilon_{\textbf{MA}}}$ and $\bm{c_i}$ denotes the codeword assigned to the $i$-th user. Through the interference channel, the received signal $\bm{y_i} \in \mathbb{C}^{M_t \times 1}$ can be represented as
\begin{equation}
    \bm{y_i} = \sum_{j=1}^{N} \sqrt{P} \bm{H_{ji}} \bm{z_j} + \bm{n_i},
    \label{eq3}
\end{equation}
where $\bm{H_{ji}} \in \mathbb{C}^{M_t \times M_t}$ is the channel matrix from TX $j$ to RX $i$, $\bm{n_i} \sim \mathcal{CN}(0, \sigma_n^2 \bm{I_{M_t}})$ is the additive white Gaussian noise (AWGN) at RX $i$ and $P$ is the transmit power of TX.

The receiver has the symmetrical structure with the transmitter. At the $i$-th RX, the received signal $\bm{y_i}$ is processed through a multi-user detection module to extract the semantic information $\bm{\hat{x}_i} \in \mathbb{R}^{M_t \times 1}$ and ultimately reconstruct the input image $\bm{\hat{s}_i} \in \mathbb{R}^{H \times W \times 3}$ throuth the semantic decoder. The process can be formally expressed as:
\begin{align}
    \bm{\hat{x}_i} & = g_{\text{MUD}}(\bm{y_i}, \, \bm{c_i} \,;\, \bm{\zeta_{\textbf{MUD}}}), \label{eq5} \\
    \bm{\hat{s}_i} & = f_{\text{dec}}(\bm{\hat{x}_i} \,;\, \bm{\Phi_{\textbf{dec}}}), \label{eq6}
\end{align}
where $g_{\text{MUD}}(\, \bm{\cdot} \,;\, \bm{\zeta_{\textbf{MUD}}} \,)$ denotes the multi-user detection mapping function, parameterized by $\bm{\zeta_{\textbf{MUD}}}$, and $f_{\text{dec}}(\, \bm{\cdot} \,;\, \bm{\Phi_{\textbf{dec}}} \,)$ denotes the semantic decoding function parameterized by $\bm{\Phi_{\textbf{dec}}}$. 

To maximize the reconstruction quality within the multi-user interference system model, the training loss function can be defined as follows:
\begin{equation}
    \mathop{\arg\min}_{\bm{\Theta}, \, \bm{\varepsilon}, \, \bm{\zeta}, \, \bm{\Phi}} \sum_{i=1}^{N} \mathbb{E}_{\bm{s_i} \sim p_{\bm{s_i}}} \mathbb{E}_{\bm{\hat{s}_i} \sim p_{\bm{\hat{s}_i} \mid \bm{s_i}}} [d( \bm{s_i}, \bm{\hat{s}_i} )],
    \label{eq7}
\end{equation}
where $\bm{\Theta}, \, \bm{\varepsilon}, \, \bm{\zeta}, \, \bm{\Phi}$ encapsulate all the network parameters with subscripts omitted for simplicity, $\bm{s_i}$ and $\bm{\hat{s}_i}$ refers to the input raw RGB images and the reconstructed images, respectively, and $d( \bm{s_i}, \bm{\hat{s}_i} )$ denotes the distortion measurement of the original and reconstructed images. 



\subsection{NOC-Based Multi-User Access Mechanism}

We propose a novel multi-user access mechanism based on NOCs to mitigate inter-user interference without bandwidth expansion. NOCs are specifically designed to facilitate multi-user access by mapping user-specific semantic information into distinguishable feature subspaces. Unlike CDMA, NOCs intentionally relax the strict orthogonality, allowing for increased flexibility without additional bandwidth expansion. In this paper, each TX and RX pair employs the NOC-based multi-user access mechanism to mitigate inter-user interference. 

Unlike existing studies that rely on user- or task-specific architectures requiring individualized encoder–decoder parameters, which substantially increase training parameters and computational costs, this paper employs NOCs to map the semantic information $\bm{x}$ of different users to distinct semantic feature subspaces, while maintaining identical NN parameters for each user. Let $\mathcal{X}$ and $\mathcal{F}_i$ denotes the semantic information space and the semantic feature subspace associated with user $i$, which can be defined as follows:
\begin{align}
    \mathcal{X} & = \{ \bm{x_i} \in \mathbb{R}^{M_t \times 1} \mid \bm{x_i} = f_{\text{enc}}(\bm{s_i} \, ; \, \bm{\Theta}), \, \forall \bm{s_i} \in \mathcal{D} \}, \label{eq8} \\
    \mathcal{F}_i & = \{ \bm{z_i} \in \mathbb{C}^{M_t \times 1} \mid \bm{z}_i = g_{\text{MA}}(\bm{x_i}, \, \bm{c_i} \,;\, \bm{\varepsilon}), \, \forall \bm{x}_i \in \mathcal{X} \}, \label{eq9}
\end{align}
where $\bm{\Theta}$ and $\bm{\varepsilon}$ denotes the parameters with subscripts omitted for simplicity, and $\mathcal{D}$ is the knowledge base of the source. 

Given the sharing of NN parameters and the uniformity of knowledge backgrounds across all users, the semantic information $\bm{x}$ are situated within the same semantic information space $\mathcal{X}$. To mitigate inter-user interference, the semantic feature subspace $\mathcal{F}_i$ must satisfy the following conditions:

\textit{Subspace Separation:} Each user's feature $\bm{z} \in \mathbb{C}^{M_t \times 1}$ should reside in its designated semantic feature subspace, ensuring mutual exclusivity:
\begin{subequations}
    \begin{align}
        & \bm{z_i} \in \mathcal{F}_i, \:\: \bm{z_i} \notin \mathcal{F}_j, \quad \forall i \neq j, \label{eq10a} \\ 
        & \mathcal{F}_i \cap \mathcal{F}_j = \emptyset, \quad \forall i \neq j.
        \label{eq10b}
    \end{align}
\end{subequations}

\textit{Feature Orthogonality:} The semantic feature $\bm{z} \in \mathbb{C}^{M_t \times 1}$ of different users should achieve orthogonality, donated as:
\begin{equation}
    \langle \bm{z_i}, \bm{z_j} \rangle = 0, \quad \forall i \neq j, \: \bm{z_i} \in \mathcal{F}_i, \:  \bm{z_j} \in \mathcal{F}_j,
    \label{eq11}
\end{equation}
where $\langle \cdot \rangle$ denotes the inner product operation.

\textit{Interference Power Constraint:} The projection power of each user's semantic feature $\bm{z}$ in other users' subspaces should be minimized to reduce interference:
\begin{equation}
    \left\| \text{Proj}_{\mathcal{F}_j}(\bm{z_i}) \right\|^2 \leq \epsilon_0, \quad \forall i \neq j,
    \label{eq12}
\end{equation}
where $\text{Proj}_{\mathcal{F}_j}(z_i)$ denotes the projection of $z_i$ onto the subspace $\mathcal{F}_j$ and $\epsilon_0$ is a small positive constant.

\begin{figure*}[t]
\centering
\includegraphics[width=0.95\textwidth]{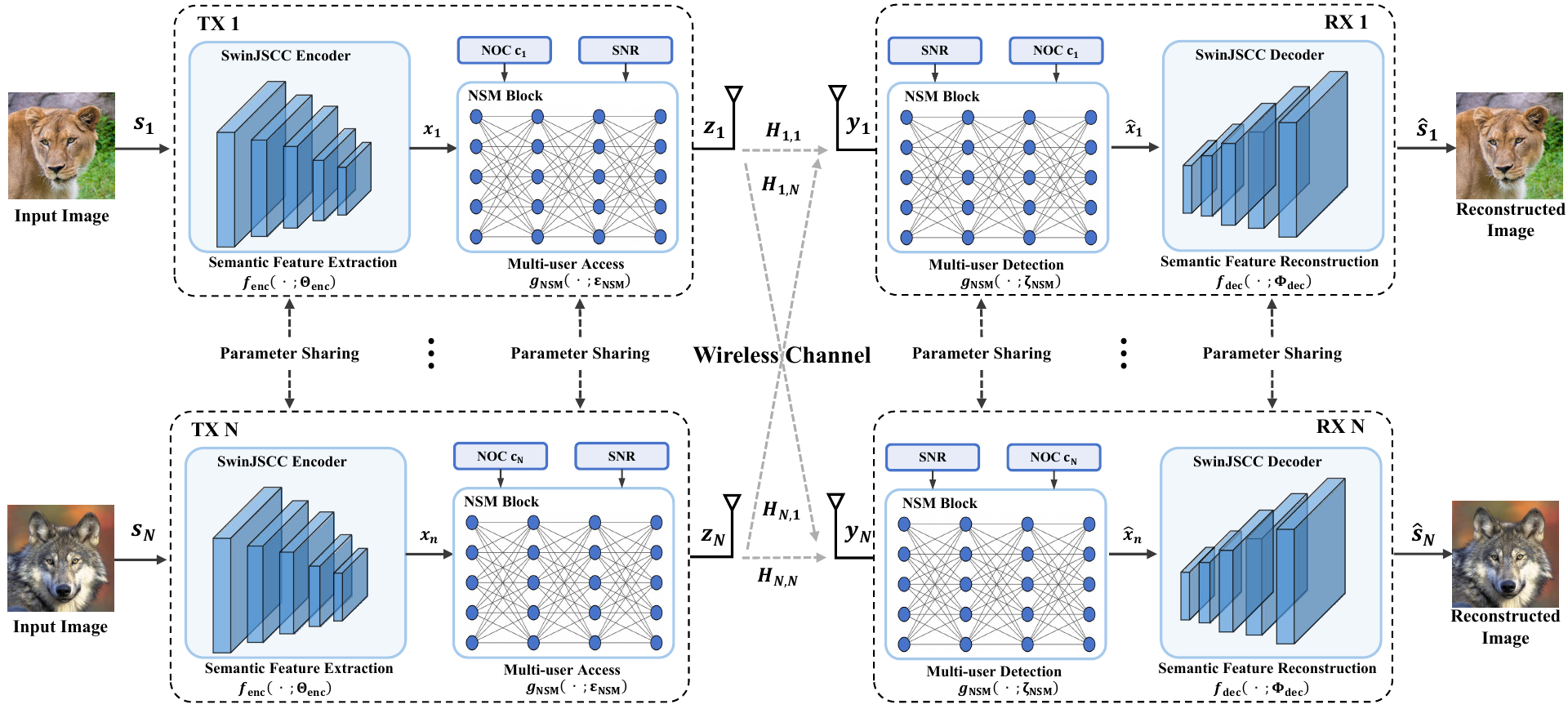}
\caption{The overall architecture of multi-user semantic communication system with non-orthogonal codewords for image transmission.}
\label{fig2}
\end{figure*}

Then, through the interference channel, the received signal $\bm{y_i}$ at RX $i$ can be expressed as
\begin{subequations}
    \begin{align}
        \bm{y_i} = & \quad \bm{H_{ii}} \, \bm{z_i} + \sum_{j=1, \, j \neq i}^{N} \bm{H_{ji}} \, \bm{z_j} + \bm{n_i} \\
                 = & \quad \underbrace{\bm{H_{ii}} \;  g_{\text{MA}}(\bm{x_i}, \, \bm{c_i} \,;\, \bm{\varepsilon_{\textbf{MA}}})}_{\text{Disired signal subspace}} \notag \\
                   & + \underbrace{\sum_{j=1, \; j \neq i}^{N} \bm{H_{ji}} \; g_{\text{MA}}(\bm{x_j}, \, \bm{c_j} \,;\, \bm{\varepsilon_{\textbf{MA}}})}_{\text{Interference subspace}} + \underbrace{\bm{n_i}}_{\text{Noise}}
        \label{eq13}
    \end{align}
\end{subequations}
where $\bm{c_i}$ assigned to the user $i$ is a $N_t \times 1$ column vector and each element $c_j \in \{-1, +1\}$ for $j = 1, 2, \ldots, M_t$. The received signal $\bm{y_i}$ consisting of three main components: the disired signal subspace, the interference subspace and noise.

Moreover, the received interference signal $\bm{y_i}$ is fed into the multi-user detection module at RX $i$ to extract the disired signal as follows:
\begin{subequations}
\begin{align}
    \bm{\hat{x}_i} = & \: g_{\text{MUD}}(\bm{y_i}, \, \bm{c_i} \,;\, \bm{\zeta_{\textbf{MUD}}}) \\
             = & \: g_{\text{MUD}}(\bm{H_{ii}} \, \bm{z_i} + \sum_{j=1, \, j \neq i}^{N} \bm{H_{ji}} \, \bm{z_j} + \bm{n_i}, \, \bm{c_i} \, ;\, \bm{\zeta_{\textbf{MUD}}}),
    \label{eq14}
\end{align}
\end{subequations}
where the interference components $\sum_{j=1, \, j \neq i}^{N} \bm{H_{ji}} \bm{z_j}$ are projected into the semantic feature subspace $\mathcal{F}_i$ of user $i$ using $\bm{c_i}$ through the function $g_{\text{MUD}}(\,\bm{\cdot}\,;\, \bm{\zeta_{\textbf{MUD}}})$. Leveraging the Feature Orthogonality and the Interference Power Constraint, the resulting projected values are effectively negligible, thus eliminating inter-user interference and enabling the reconstruction of the desired original images through the semantic decoder. Furthermore, since each user’s semantic features are mapped to a unique subspace determined by its assigned codeword, meaningful decoding is only possible with the correct codeword, thereby inherently preventing unauthorized reconstruction by other users.

\textbf{Noting:} The Subspace Separation and Feature Orthogonality conditions can be effectively enforced during training by introducing corresponding loss functions, ensuring that the learned semantic feature subspaces remain mutually exclusive and approximately orthogonal across users.

\section{Multi-User Semantic Communication System with Non-Orthogonal Codewords}

In this section, we present multi-user SemCom system with NOCs system for wireless image transmission. Inspired by \cite{ref40}, we employ the SwinJSCC architecture as both the semantic encoder and decoder. Moreover, we introduce a novel NOCs and SNR modulation block named NSM block that functions as both the multi-user access and multi-user detection. Finally, we design a fixed-angle NOCs construction scheme based on the Walsh matrix.

\begin{figure*}[t]
    \centering
    \begin{minipage}{0.63\textwidth}
        \subfloat[\textbf{(a) SwinJSCC Encoder}]{\includegraphics[width=0.98\linewidth]{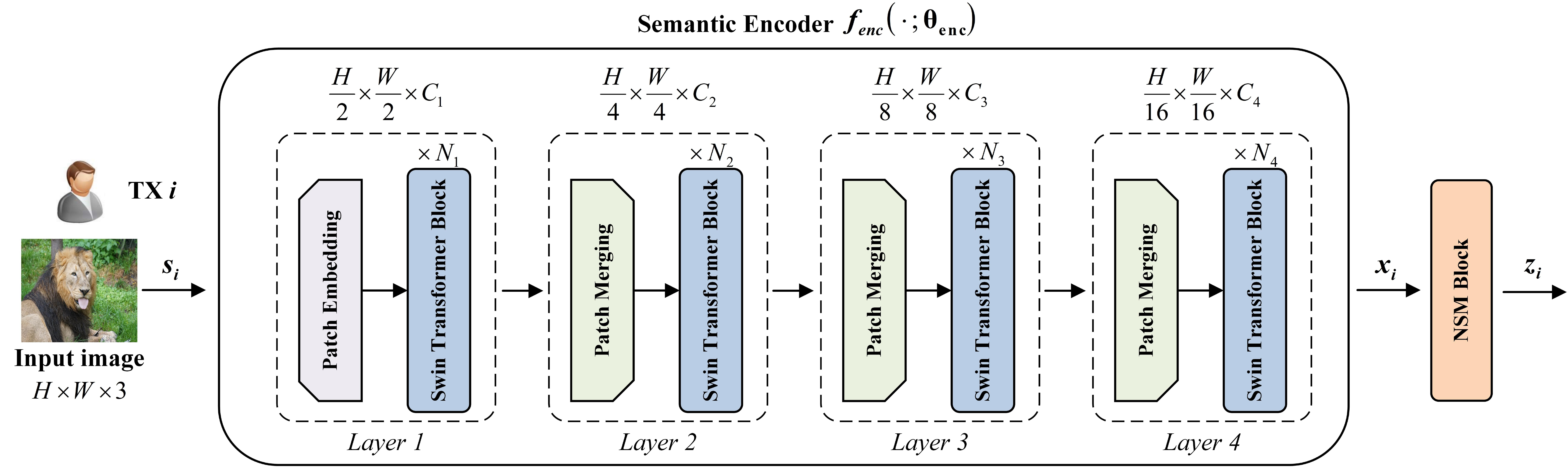}%
        \label{fig3a}}
        \hfill
        \subfloat[\textbf{(b) SwinJSCC Decoder}]{\includegraphics[width=0.98\linewidth]{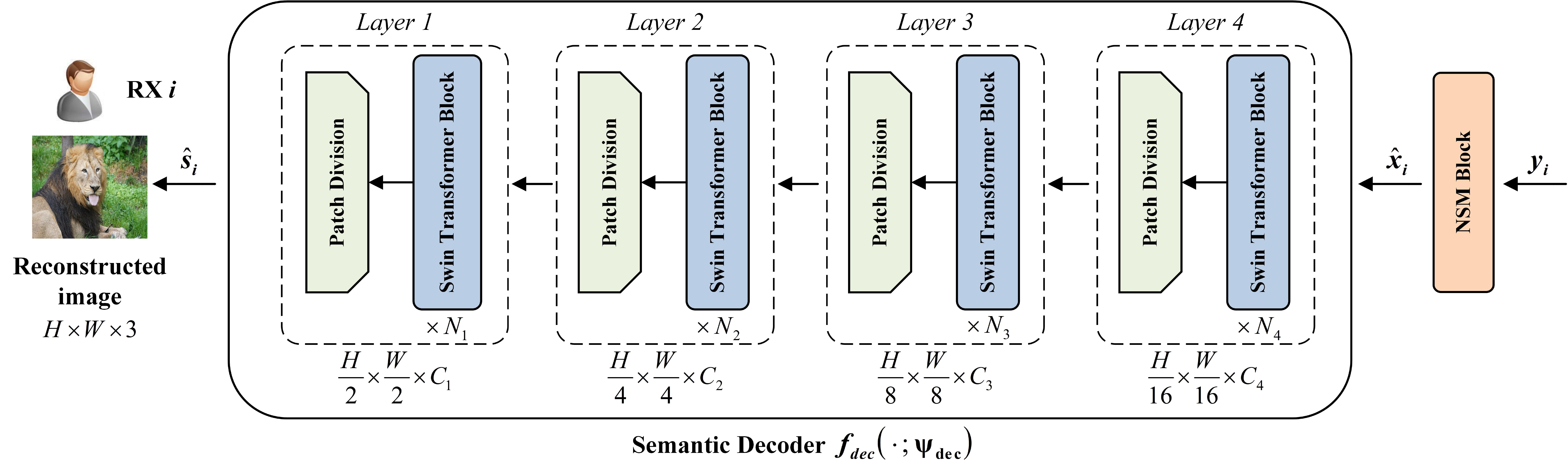}%
        \label{fig3b}}
    \end{minipage}
    \begin{minipage}{0.33\textwidth}
        \centering
        \subfloat[\textbf{(c) The architecture of NSM Block}]{\includegraphics[width=0.87\linewidth]{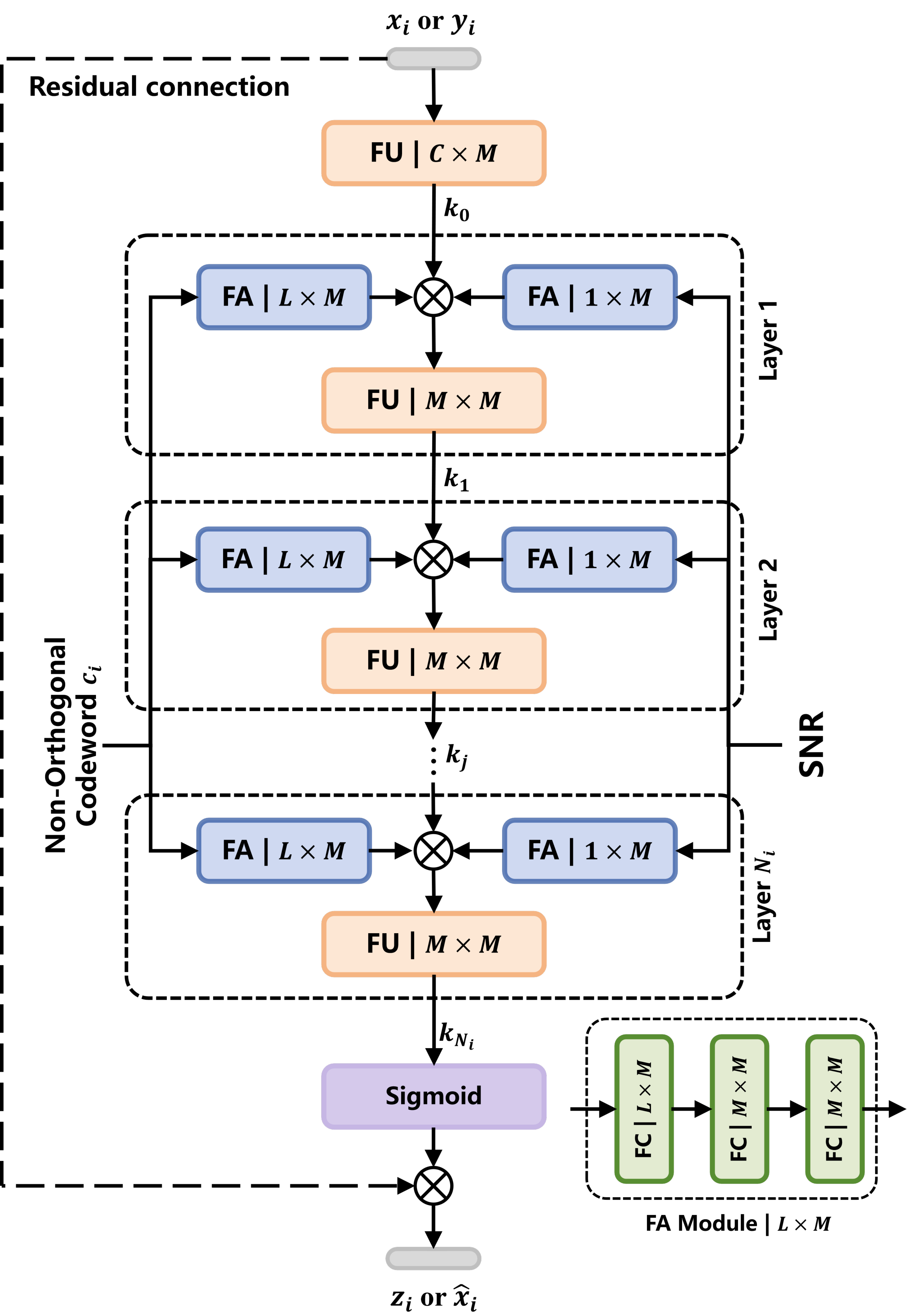}%
        \label{fig3c}}
    \end{minipage}
    \caption{(a) The overall architecture of the SwinJSCC Encoder for extracting semantic information from images. (b) The overall architecture of the SwinJSCC Decoder for reconstructing images from the semantic information. (c) The architecture of NSM Block.}
    \label{fig3}
\end{figure*}

\subsection{The Overall Architecture of NOC4SC}

The overall system framework of NOC4SC is depicted in Fig.\ref{fig2}, where a scenario with $N$ users is considered to illustrate the system’s functionality. In the semantic encoder of TX $i$ shown in Fig.\ref{fig3}(a), the Patch Embedding (PE) layer $l_{\text{PE}}(\, \cdot \,; \, \bm{\Theta_{\textbf{PE}}})$ partitions the input RGB images $\bm{s_i} \in \mathbb{R}^{H \times W \times 3}$ into non-overlapping patches of size $\frac{H}{2} \times \frac{W}{2}$. Then the Swin Transformer (ST) Block $l_{\text{ST}}(\, \cdot \,; \, \bm{\Theta_{\textbf{ST}}})$ processes the features while preserving the same resolution. This sequence of operations can be expressed as
\begin{equation}
    L_{1}(\, \cdot \,; \, \bm{\Theta_{\textbf{L}_{1}}}) =  l_{\text{ST}_{N_{1}}}(\, \cdots l_{\text{ST}_1}(\, l_{\text{PE}}(\, \cdot \,) \,) \cdots \,),
    \label{eq15}
\end{equation}
where $\bm{\Theta_{\textbf{L}_{1}}} = \{\bm{\Theta_{\textbf{PE}}}, \, \bm{\Theta_{\textbf{ST}_{1}}}, \, \cdots, \, \bm{\Theta_{\textbf{ST}_{N_{1}}}}\}$ represents the 
network parameters of Layer 1.

Futhermore, to extract deeper semantic information representations, the Patch Merging (PM) layer $l_{\text{PM}}(\, \cdot \,; \, \bm{\Theta_{\textbf{PM}}})$ aggregates adjacent features and performs down-sampling through linear layers, reducing the token count by half \cite{ref41}. The sequence of operations from the PM layer to the $N_j$ ST Block defines "$Layer \: j$", expressed as
\begin{equation}
    L_{j}(\, \cdot \,; \, \bm{\Theta_{\textbf{L}_j}}) =  l_{\text{ST}_{N_{j}}}(\, \cdots l_{\text{ST}_1}(\, l_{\text{PM}}(\, \cdot \,) \,) \cdots \,),
    \label{eq16}
\end{equation}
where $\bm{\Theta_{\textbf{L}_j}} = \{\bm{\Theta_{\textbf{PM}}}, \, \bm{\Theta_{\textbf{ST}_{1}}}, \, \cdots, \, \bm{\Theta_{\textbf{ST}_{N_{j}}}}\}$ for $1 < j \leq N_i$ represents the parameters of Layer $j$. Notably, it is necessary to select an appropriate number of layers in the semantic encoder to accommodate different image resolutions.


In summary, the output through the SwinJSCC Encoder in Tx $i$ is as follows:
\begin{equation}
    \bm{x_i} = f_{\text{enc}}(\bm{s_i} \, ; \, \bm{\Theta_{\textbf{enc}}}),
    \label{eq19}
\end{equation}
\begin{equation}
    f_{\text{enc}}(\, \cdot \, ; \, \bm{\Theta_{\textbf{enc}}}) =  L_{N}(\, \cdots \, L_{1}(\, \cdot \,; \, \bm{\Theta_{\textbf{L}_1}}) \, \cdots \,; \, \bm{\Theta_{\textbf{L}_N}}),
    \label{eq20}
\end{equation}
where $f_{\text{enc}}(\, \cdot \, ; \, \bm{\Theta_{\textbf{enc}}})$ refers to the SwinJSCC encoder function with the network parameters $\bm{\Theta_{\textbf{enc}}} = \{\bm{\Theta_{\textbf{L}_1}}, \, \bm{\Theta_{\textbf{L}_2}}, \, \cdots, \, \bm{\Theta_{\textbf{L}_N}}\}$.

Then, the semantic information $\bm{x_i}$ is modulated into the semantic feature $\bm{z_i}$ by NSM Block with SNR and NOC $\bm{c_i}$ assigned to the TX $i$ as follows:
\begin{subequations}
    \begin{align}
        \bm{z_i} = & \: g_{\text{NSM}}(\bm{x_i}, \, \bm{c_i} \,;\, \bm{\varepsilon_{\textbf{NSM}}}) \\
                 = & \: g_{\text{NSM}}(f_{\text{enc}}(\bm{s_i} \, ; \, \bm{\Theta_{\textbf{enc}}}), \, \bm{c_i} \,;\, \bm{\varepsilon_{\textbf{NSM}}})
        \label{eq21}
    \end{align}
\end{subequations}
where $g_{\text{NSM}}(\, \bm{\cdot} \,;\, \bm{\varepsilon_{\textbf{NSM}}} \,)$ represents the NSM Block mapping function, parameterized by $\bm{\varepsilon_{\textbf{NSM}}}$. The detailed discussion of NSM Block will be elaborated upon in Section III-B. 

As shown in Fig.\ref{fig1}, $N$ TXs simultaneously send semantic feature vectors $\bm{z_i}$ to N RXs, leading to inevitable inter-user interference and noise within the wireless channel. The structure of the RX $i$ is symmetrical to the TX $i$, as illustrated in Fig.\ref{fig3}(b). The interference signal $\bm{y_i}$ is processed through the NSM Block to isolate the desired signal, as follows:
\begin{equation}
    \bm{\hat{x}_i} =  g_{\text{NSM}}(\bm{H_{ii}} \, \bm{z_i} + \sum_{j=1, \, j \neq i}^{N} \bm{H_{ji}} \, \bm{z_j} + \bm{n_i}, \, \bm{c_i} \, ;\, \bm{\zeta_{\textbf{NSM}}}),
    \label{eq23}
\end{equation}
where the NOC $\bm{c_i}$ at the RX $i$ should correspond to the codeword assigned at the TX $i$ and $g_{\text{NSM}}(\, \bm{\cdot} \,;\, \bm{\zeta_{\textbf{NSM}}} \,)$ represents the NSM Block mapping function, parameterized by $\bm{\zeta_{\textbf{NSM}}}$.

\begin{algorithm}[t]
    \renewcommand{\algorithmicrequire}{\textbf{Input:}}
    \renewcommand{\algorithmicensure}{\textbf{Output:}}
    \caption{Training Process for NOC4SC System}
    \label{alg:alg1}
    \begin{algorithmic}[1]
    
        \REQUIRE learning rate $\eta$,  epochs $E$, number of users $N$, and non-orthogonal codewords $\{\bm{c_j}\}_{j=1}^{N}$.
        
        \ENSURE Optimized parameters $\{\bm{\Theta^*}, \, \bm{\varepsilon^*}, \, \bm{\zeta^*}, \, \bm{\Phi^*}\}$.
        
        \STATE \textbf{Initialize:} Initialized network parameters $\{\bm{\Theta}, \, \bm{\varepsilon}, \, \bm{\zeta}, \, \bm{\Phi}\}$.
        
        \FOR{$e = 1, 2, \dots, E$}
            \FOR{$i = 1, 2, \dots, N$}
                \STATE Resample from dataset $\mathcal{D} \rightarrow \bm{s_i}$;
                \STATE Abstract semantic information by (\ref{eq19});
                \STATE Calculate semantic feature by (\ref{eq21});
            \ENDFOR
            \STATE Generate channel gain and noise
            \[
            \begin{aligned}
                \bm{H}_{ij} &\sim \mathcal{CN}(0, \sigma_n^2 \bm{I}_{M_t}), \quad i,\, j \in \{1,\dots,N\}; \\
                \bm{n}_{i} &\sim \mathcal{CN}(0, \sigma_n^2 \bm{I}_{M_t});
            \end{aligned}
            \]
            \FOR{$i = 1, 2, \dots, N$}
                \STATE Received signal $\bm{y_i}$ by (\ref{eq3});
                \STATE Eliminate interference by (\ref{eq23});
                \STATE Reconstructed image by (\ref{eq24});
            \ENDFOR 
            \STATE Compute the overall loss by (\ref{eq27}), (\ref{eq28}), (\ref{eq29}), (\ref{eq:orth}): 
            \[
            \mathcal{L}_\text{NOC4SC} =  \mathcal{L}_\text{recon} + \lambda_{\text{fair}} \, \mathcal{L}_{\text{fair}} + \lambda_{\text{orth}} \, \mathcal{L}_{\text{orth}};
            \]
            \STATE Update the network parameters:
            \[
            \{\bm{\Theta}, \, \bm{\varepsilon}, \, \bm{\zeta}, \, \bm{\Phi}\} \leftarrow \{\bm{\Theta}, \, \bm{\varepsilon}, \, \bm{\zeta}, \, \bm{\Phi}\} - \eta \nabla \bm{L_{\text{total}}};
            \]
        \ENDFOR 
        \RETURN The parameters $\{\bm{\Theta^*}, \, \bm{\varepsilon^*}, \, \bm{\zeta^*}, \, \bm{\Phi^*}\}$.
    \end{algorithmic}
    \label{alg1}
\end{algorithm}

Within the semantic decoder, the Patch Division (PD) module utilizes the linear layer to up-sample the features at twice the original resolution. Similarly, the semantic information $\bm{\hat{x}_i}$ is processed through the SwinJSCC Decoder to reconstruct the image. The reconstructed image is given by
\begin{equation}
    \bm{\hat{s}_i} = f_{\text{dec}}(\bm{\hat{x}_i} \, ; \, \bm{\Phi_{\textbf{dec}}}),
    \label{eq24}
\end{equation}
\begin{equation}
    f_{\text{dec}}(\, \cdot \, ; \, \bm{\Phi_{\textbf{dec}}}) =  L_{N_i}(\, \cdots \, L_{1}(\, \cdot \,; \, \bm{\Phi_{\textbf{L}_1}}) \, \cdots \,; \, \bm{\Phi_{\textbf{L}_{N_i}}}),
    \label{eq25}
\end{equation}
where $\bm{\Phi_{\text{dec}}} = \{\bm{\Phi_{\textbf{L}_1}}, \, \bm{\Phi_{\textbf{L}_2}}, \, \cdots, \, \bm{\Phi_{\textbf{L}_{N_i}}}\}$, $f_{\text{dec}}(\cdot; \, \bm{\Phi_{\textbf{dec}}})$ denotes the SwinJSCC decoder function with the parameters $\bm{\Phi_{\textbf{dec}}}$.

To effectively train the NOC4SC system, we have designed a tailored loss function that differs from (\ref{eq7}). The optimized parameters are denoted as $\{\bm{\Theta_{\textbf{enc}}}, \, \bm{\varepsilon_{\textbf{NSM}}}, \, \bm{\zeta_{\textbf{NSM}}}, \, \bm{\Phi_{\textbf{dec}}}\}$. The overall loss function is given by:
\begin{equation}
    \mathcal{L}_\text{NOC4SC} =  \mathcal{L}_\text{recon} + \lambda_{\text{fair}} \, \mathcal{L}_{\text{fair}} + \lambda_{\text{orth}} \, \mathcal{L}_{\text{orth}},
    \label{eq27}
\end{equation}
\begin{equation}
    \mathcal{L}_\text{recon} = \frac{1}{N} \sum_{i=1}^{N} \textbf{MSE}(\bm{s_i}, \, \bm{\hat{s}_i}),
    \label{eq28}
\end{equation}
\begin{equation}
    \mathcal{L}_\text{fair} = \frac{1}{N} \sum_{i=1}^{N} \: \lvert \textbf{MSE}(\bm{s_i}, \, \bm{\hat{s}_i}) - \mathcal{L}_\text{recon} \rvert,
    \label{eq29}
\end{equation}
\begin{equation}
    \mathcal{L}_\text{orth} = \frac{1}{\binom{N}{2}} \sum_{1 \le i < j \le N} \: (\frac{\bm{z_i}}{||\bm{z_i}||_2} \cdot \frac{\bm{z_j}}{||\bm{z_j}||_2})^2,
    \label{eq:orth}
\end{equation}
where $\lambda_{\text{fair}}$ and $\lambda_{\text{orth}}$ are predetermined hyperparameters, $\binom{N}{2} = \frac{N!}{2!(N-2)!}$ represents the binomial coefficient, $\mathcal{L}_\text{recon}$ denotes each user's image reconstruction loss, $\mathcal{L}_\text{fair}$ is intended to ensure that MSE calculated for each user is approximately equal, thereby promoting fairness across all users, and $\mathcal{L}_{\text{orth}}$ denotes the average squared cosine similarity between different user features $\bm{z_i}$, penalizing cross-user correlation.

In \eqref{eq10a}–\eqref{eq11}, subspace separability and inter-user orthogonality are determined by the cross-user projections of the post-NSM Block features. The orthogonality loss $\mathcal{L}_{\text{orth}}$ acts as an empirical surrogate, where a smaller $\mathcal{L}_{\text{orth}}$ reflects stronger orthogonality, consequently reducing cross-user projection.

The whole trainging process is shown in Algorithm \ref{alg:alg1}.

\subsection{NSM Block}

In this subsection, we propose a NOC-based multi-user access modulation mechanism utilizing NOC $\bm{c_i}$, along with an adaptive modulation technique with channel state SNR, termed as the NSM Block. In Tx $i$, the semantic information $\bm{x_i}$ is modulated to the semantic feature $\bm{z_i}$ by the NSM Block incorporating NOC $\bm{c_i}$ assigned for $i$-th user and channel state SNR. In RX $i$, the received signal $\bm{y_i}$ is similarly modulated to recover the semantic information $\bm{\hat{x}_i}$ from inter-user interference and noise. \textit{\textbf{Importantly, NSM Block induces soft separability in the feature domain without any spectrum spreading}}.

The NSM Block is seamlessly embedded within both the encoder and the decoder, comprising two principal components: the Feature Alignment (FA) module and the Feature Fusion (FU) module, as illustrated in Fig.\ref{fig3}(c). The FA module is structured as a sequence of three fully connected (FC) layers, where $L$ indicates the dimensionality of the NOC $\bm{c_i}$ and $M$ represents the dimensionality of the latent vectors. The FU module comprises a single fully connected layer, where $C$ denotes the number of channels in the input vector. Concretely, NSM Block proceeds in two steps:

\textbf{Step 1 (Feature Alignment).}
The input feature $ \in \mathbb{R}^{K\times C}$ (i.e., semantic information $\bm{x_i}$ at Tx or received signal $\bm{y_i}$ at Rx) is first projected from $C$ to $M$ channels:
\begin{equation}
    \bm{k_0} = \mathbf{FU}_{0}(\bm{\text{input}}), \quad \bm{k_0} \in \mathbb{R}^{K \times M}.
    \label{eq30}
\end{equation}

In parallel, the NOC $\bm{c_i} \in \mathbb{R}^{K \times L}$ and $\textbf{SNR} \in \mathbb{R}^{K \times 1}$ are mapped by the two FA branches, which are combined into gates for subsequent fusion:
\begin{align}
    \bm{a_j} & = \mathbf{FA_{MA_j}}(\bm{c_i}), \quad \bm{a_j} \in \mathbb{R}^{K \times M}, \label{eq31a} \\
    \bm{b_j} & = \mathbf{FA_{s_j}}(\mathbf{SNR}), \quad \bm{b_j} \in \mathbb{R}^{K \times M}, \label{eq31b}
\end{align}
where $\mathbf{FA}_{\textbf{S}_{\bm{j}}}(\, \cdot \,)$ and $\mathbf{FA}_{\textbf{MA}_{\bm{j}}}(\, \cdot \,)$ denote the mapping function of $\textbf{SNR}$ and $\bm{c_i}$, respectively. This step aligns the channel width (from $C$ to $M$) and conditions the features using identity and channel state, preparing layer-wise gates for fusion.

\textbf{Step 2 (Feature Fusion).}
For each layer $j=1,\dots,N_t$, the candidate update produced by $\mathbf{FA}_{j}$ is modulated:
\begin{align}
    \bm{g_j} & = \bm{k_{j-1}} \odot \bm{a_j} \odot \bm{b_j} \quad \bm{g_j} \in \mathbb{R}^{K \times M}, \label{eq32a} \\
    \bm{k_j} & = \mathbf{FU_j}(\bm{g_i}), \quad\quad \bm{k_j} \in \mathbb{R}^{K \times M}, \label{eq32b}
\end{align}
where $\bm{k_j}$ represents the latent features produced by the $\mathbf{FU}_{j}$ layer, and $\odot$ is the Hadamard product \cite{ref42}. After $N_t$ layers, the final gate is applied to the input to obtain the block output:
\begin{equation}
    \textbf{output} = \textbf{input} \odot \textbf{Sigmoid}(\bm{k_{N_i}}),
    \label{eq32}
\end{equation}
where $\textbf{output}$ represents the semantic feature $\bm{z_i}$ in Tx or the recovered semantic information $\bm{\hat{x}_i}$ in Rx.

\textbf{Noting:} The above two-step procedure implements identity- and SNR-aware modulation in the latent space, yielding soft separability among users without spectrum spreading or power-domain SIC.

\subsection{Design of Non-Orthogonal Codewords}

To facilitate multi-user access, we design a fixed-angle NOCs construction scheme by perturbing orthogonal codes derived from the Walsh matrix. The Walsh matrix is an orthogonal code structure extensively used in classical CDMA systems due to its binary format and ideal cross-correlation properties. The Walsh matrix, denoted as $\bm{W_{\textbf{N}}}$, is an $N \times N$ matrix generated recursively using the Hadamard construction:
\begin{equation}
    \bm{W_{1}} = \begin{bmatrix} 1 \end{bmatrix}, \quad
    \bm{W_{2\textbf{N}}} = \begin{bmatrix} \bm{W_{\textbf{N}}} & \bm{W_{\textbf{N}}} \\ \bm{W_{\textbf{N}}} & -\bm{W_{\textbf{N}}} \end{bmatrix}.
    \label{bmatrix}
\end{equation}

\begin{algorithm}[t]
    \renewcommand{\algorithmicrequire}{\textbf{Input:}}
    \renewcommand{\algorithmicensure}{\textbf{Output:}}
    \caption{Fixed-Angle NOCs Generation} 
    \label{alg:NOC_generation} 
    \begin{algorithmic}[1]
        \REQUIRE Orthogonal codebook $\bm{W} \in \mathbb{R}^{N \times N}$, number of users $K$, codeword length $N$, target angle $\theta_M$
        \ENSURE Non-orthogonal codebook $\{\bm{c_1}, \dots, \bm{c_K} \}$
        \STATE Compute target dot product: $d_{\text{target}}\leftarrow \text{cos}(\theta_M) \cdot N$  
        \STATE Initialize: Select $K$ rows from $\bm{W}$ as $\{\bm{c_1}, \dots, \bm{c_K} \}$
        \FOR{$\text{iteration} = 1$ to iters}
            \FOR{each pair $(i,j)$, $i<j$}
                \STATE Compute $d_{ij} \leftarrow \bm{c_i}^{\text{T}}\bm{c_j}$
                \IF{$|d_{ij} - d_{\text{target}}| > \epsilon$}
                    \FOR{$k=1$ to $N$}
                        \STATE Flip sign of $\bm{c_i}[k]$
                        \STATE Recompute $d_{ij}^{'} \leftarrow \bm{c_i}^{\text{T}}\bm{c_j}$
                        \IF{$|d_{ij}^{'} - d_{\text{target}}| < |d_{ij} - d_{\text{target}}|$}
                            \STATE Accept the flip
                        \ELSE
                            \STATE Revert the flip
                        \ENDIF
                    \ENDFOR
                \ENDIF
            \ENDFOR 
        \ENDFOR
        \RETURN $\{\bm{c_1}, \dots, \bm{c_K} \}$
    \end{algorithmic}
\end{algorithm}

Each row of the Walsh matrix is an orthogonal binary sequence, and the inner product between any two distinct rows is zero. To generate NOCs with a specified angle, we perturb the original Walsh sequences by selectively flipping the binary elements. This process transforms the orthogonal codebook into a set of codewords whose pairwise inner products closely approximate the predefined target angle. Formally, given that each codeword element is restricted to values of either $+1$ or $-1$ and the codeword length is fixed at $N$, the optimization of codewords can be rigorously formulated as follows:
\begin{equation}
    \min_{\{\bm{c_i}\}} \sum_{i \neq j}\left(\bm{c_i}^\mathrm{T}\bm{c_j} - N\cos\theta_{\text{target}}\right)^2,
    \label{eq33}
\end{equation}
where $\bm{c_i}$ is the codeword assigned to the user $i$, and $\theta_{\text{target}}$ denotes the desired angle. The algorithmic implementation is detailed in Algorithm~\ref{alg:NOC_generation}.

NOC4SC assigns each user a predefined NOC and induces soft separability in the latent space without spectrum spreading. As the number of users increases, the separation margin between feature subspaces narrows, resulting in greater feature overlap and residual inter-user interference, which in turn degrades the reconstruction quality. When the number of users exceeds the robust operating regime, NOC4SC can be composed with conventional multiple-access schemes, without altering the overall architecture.

\textbf{Noting:} In NOC4SC, each user is assigned a predefined NOC as its semantic identity. Together with the NSM module, these identities induce soft separability in the latent representation space rather than enforcing hard orthogonality at the codeword level. Fixing the pairwise angle between users does not increase the theoretical upper bound on the number of simultaneously separable users. Binding each user to a distinct NOC enables simultaneous multi-user access over shared time–frequency resources without spectrum spreading, while the parameter-shared encoder–decoder architecture ensures efficient parallel decoding.

\captionsetup[table]{justification=centering, labelsep=space, textfont=sc} 
\begin{table}[t]
    \renewcommand\arraystretch{1.8}
    \setlength{\tabcolsep}{4pt}
    \caption{ \\ Parameter Setting of NOC4SC \label{tab5}}
    \centering
    \begin{tabular}{cccc}
        \hline
        \hline
        Parameters &  & Parameters &  \\
        \hline
        Learning Rate & 1e-4 & Optimizer & Adam \\
        $\lambda_\text{fair}$ & 0.01 & $\lambda_\text{orth}$ & 0.01  \\
        NOC Angle(${}^{\circ}$) & [50, 70] & NSM Block Depth & 10  \\
        Number of Users & \multicolumn{3}{c}{[2, 3, 4, 5, 6]} \\
        BackBone & \multicolumn{3}{c}{Swin Transformer} \\
        LR Scheduler & \multicolumn{3}{c}{CosineAnnealingLR} \\
        CBR & \multicolumn{3}{c}{0.5 / 0.125 (CIFAR10 / AFHQ)} \\
        Channel Type & \multicolumn{3}{c}{AWGN, Rayleigh, Rician} \\
        \hline
        \hline
    \end{tabular}
\end{table}

\section{Experimental Results and Analysis}

\subsection{Experimental Setup}

\textit{1) Datasets:} To validate the applicability of our NOC4SC model, we utilized the CIFAR-10 \cite{ref44} dataset and AFHQ dataset \cite{choi2020stargan}. The CIFAR-10 dataset consists of 60,000 RGB images with the resolution of $32 \times 32$ pixels, providing a standard benchmark for low-resolution image reconstruction. Futhermore, the AFHQ dataset comprises 15,000 high-quality animal face images with a resolution of $512 \times 512$. During both the training and testing phases, the images of the AFHQ dataset are resized to the dimensions of $256 \times 256$.

\textit{2) Baseline Method:} To demonstrate the superiority of the proposed NOC4SC system, several other baseline methods have also been implemented for the comparison:
\begin{itemize}
    \item{\textbf{Conventional Scheme:} The conventional approach integrates JPEG image compression followed by rate-$1/2$ LDPC channel coding and BPSK digital modulation. To enable multi-user access, it adopts a classical NOMA scheme \cite{ref50} based on SCSIC, denoted as JPEG-LDPC-BPSK-NOMA.}
    \item{\textbf{OMA Scheme:} The OMA counterpart also applies JPEG compression and rate-$1/2$ LDPC channel coding, combined with 16QAM digital modulation. For fair comparison with the NOC-based design in the proposed NOC4SC system, it employs orthogonal spreading codes for CDMA-based transmission \cite{ref49}, referred to as JPEG-LDPC-16QAM-CDMA.}
    \item{\textbf{DeepJSCC Scheme:} Several DeepJSCC-based multi-user schemes are considered, including NOMASC \cite{li2023non}, DeepJSCC-NOMA \cite{yilmaz2023distributed}, and DeepJSCC-PNOMA \cite{yilmaz2025learning}. These methods employ neural JSCC to enable multi-user semantic transmission over shared channels.}
    \item{\textbf{Upper bound:} The upper bound is obtained by adopting the NOC4SC framework for point-to-point communication, where both the training and testing phases are conducted without considering inter-user interference.}
\end{itemize}

\textbf{Noting:} Since DeepJSCC-PNOMA employs a spreading matrix, its channel bandwidth rate (CBR) is proportionally reduced according to the number of users for a fair comparison. Futhermore, the semantic encoder and decoder of the NOMASC baseline are replaced with the Swin Transformer architecture, while the uperposition coding and multi-user detection mechanisms remain unchanged. Both the Conventional and OMA schemes employ perfect channel equalization.

\begin{figure}[t]
    \centering
    \includegraphics[width=0.48\textwidth]{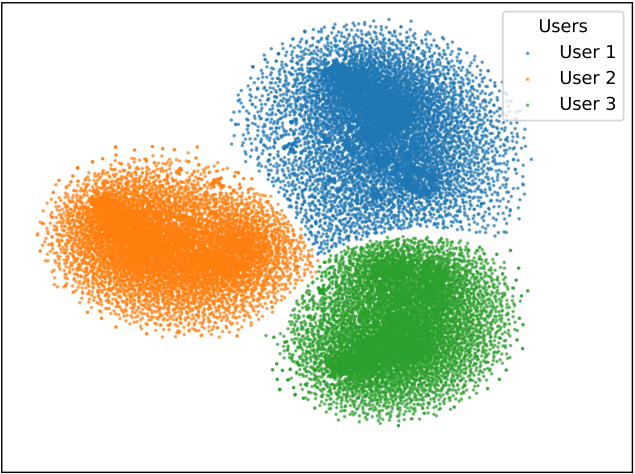}
    \caption{The t-SNE visualization of semantic feature subspaces is presented for the three-user communication on the CIFAR-10 dataset, with NOC angles set to 50° and the NSM block depth of 10.}
    \label{fig5a}
\end{figure}

\captionsetup[table]{justification=centering, labelsep=space, textfont=sc} 
\begin{table}[t]
    \renewcommand\arraystretch{2.3}
    \setlength{\tabcolsep}{8pt}
    \caption{ \\ The inner product and angle of semantic features in the three-user communication scenario\label{tab1}}
    \centering
    \begin{tabular}{cccc}
        \hline
        \hline
        Inner Product & $\frac{\bm{z_{1}}^{H} \bm{z_2}}{|\bm{z_{1}}| |\bm{z_{2}}|}$ & $\frac{\bm{z_{1}}^{H} \bm{z_3}}{|\bm{z_{1}}| |\bm{z_{3}}|}$ & $\frac{\bm{z_{2}}^{H} \bm{z_3}}{|\bm{z_{2}}| |\bm{z_{3}}|}$\\
        \hline
        Value & -0.0403 & -0.0270 & 0.0120 \\
        Angle(${}^{\circ}$) & 92.311  & 91.548 & 89.312 \\
        \hline
        \hline
    \end{tabular}
\end{table}

\textit{3) Performance Metrics:} To assess the performance of the proposed model, we adopt both semantic-level and pixel-level mertics. For semantic evaluation, the Learned Perceptual Image Patch Similarity (LPIPS) metric \cite{zhang2018unreasonable} is employed to calculate the perceptual similarity of the reconstructed images. LPIPS evaluates the perceptual difference between two images by comparing their deep feature extracted from a pretrained NN, with lower score indicates the higher perceptual similarity. For pixel-level evaluation, PSNR measures the fidelity of the reconstructed image, with higher values indicating reduced distortion. MS-SSIM evaluates perceptual similarity by considering luminance, contrast, and structural information across multiple scales, where higher values signify greater similarity.

\begin{figure*}[t]
    \centering
    \begin{minipage}{\textwidth}
        \centering
        \includegraphics[width=0.27\textwidth]{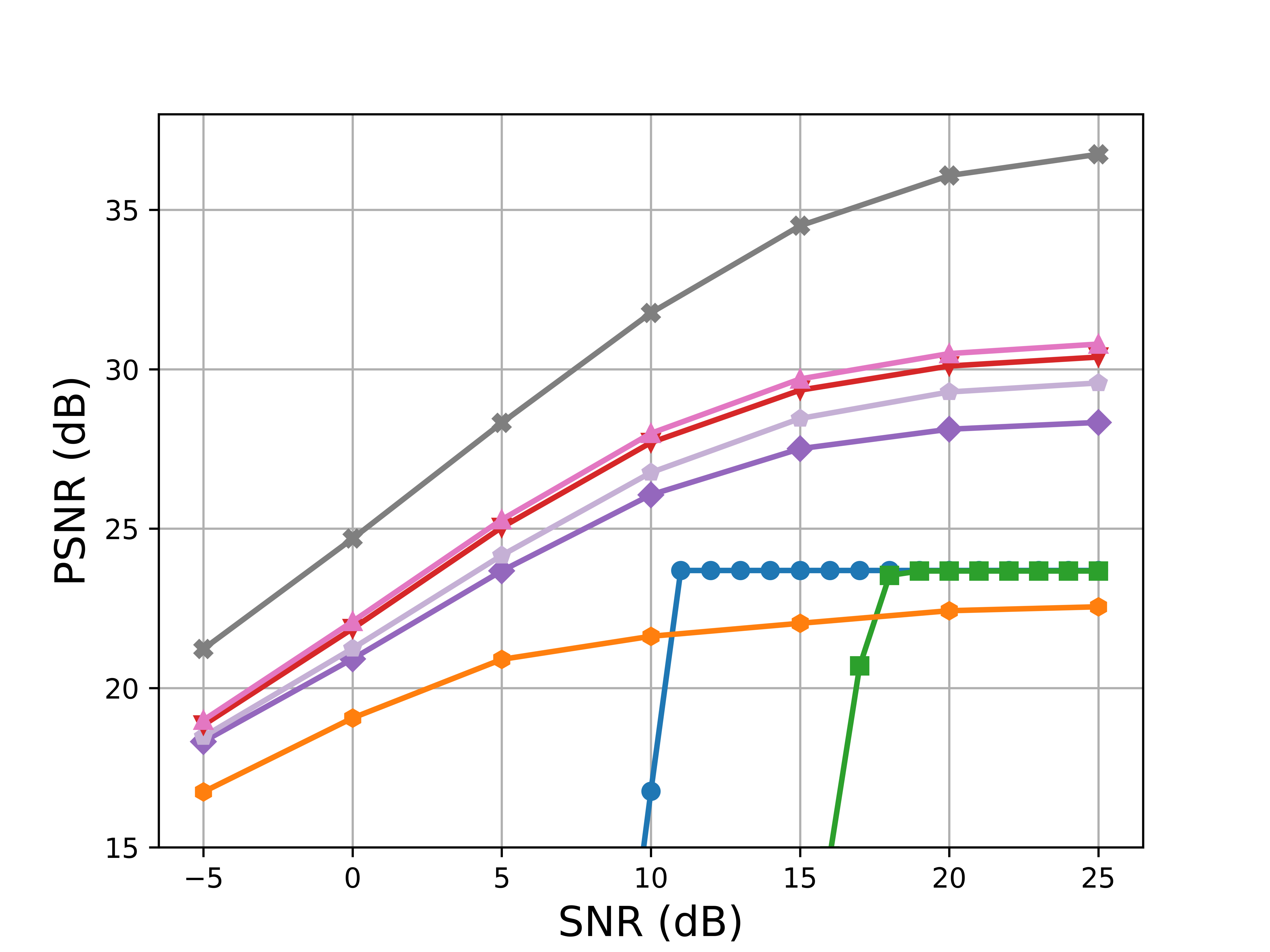} \hspace{10pt}
        \includegraphics[width=0.27\textwidth]{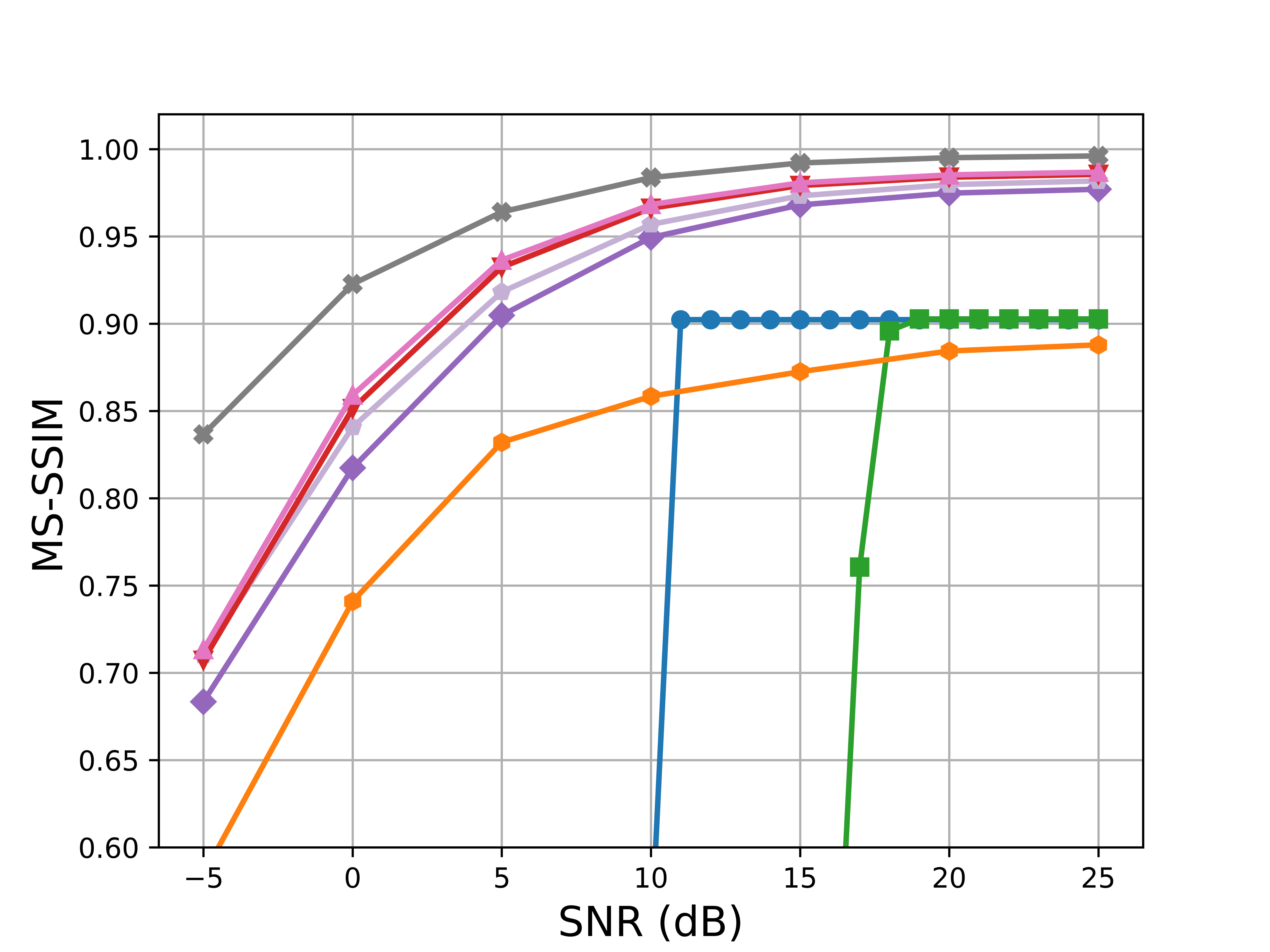} \hspace{10pt}
        \includegraphics[width=0.27\textwidth]{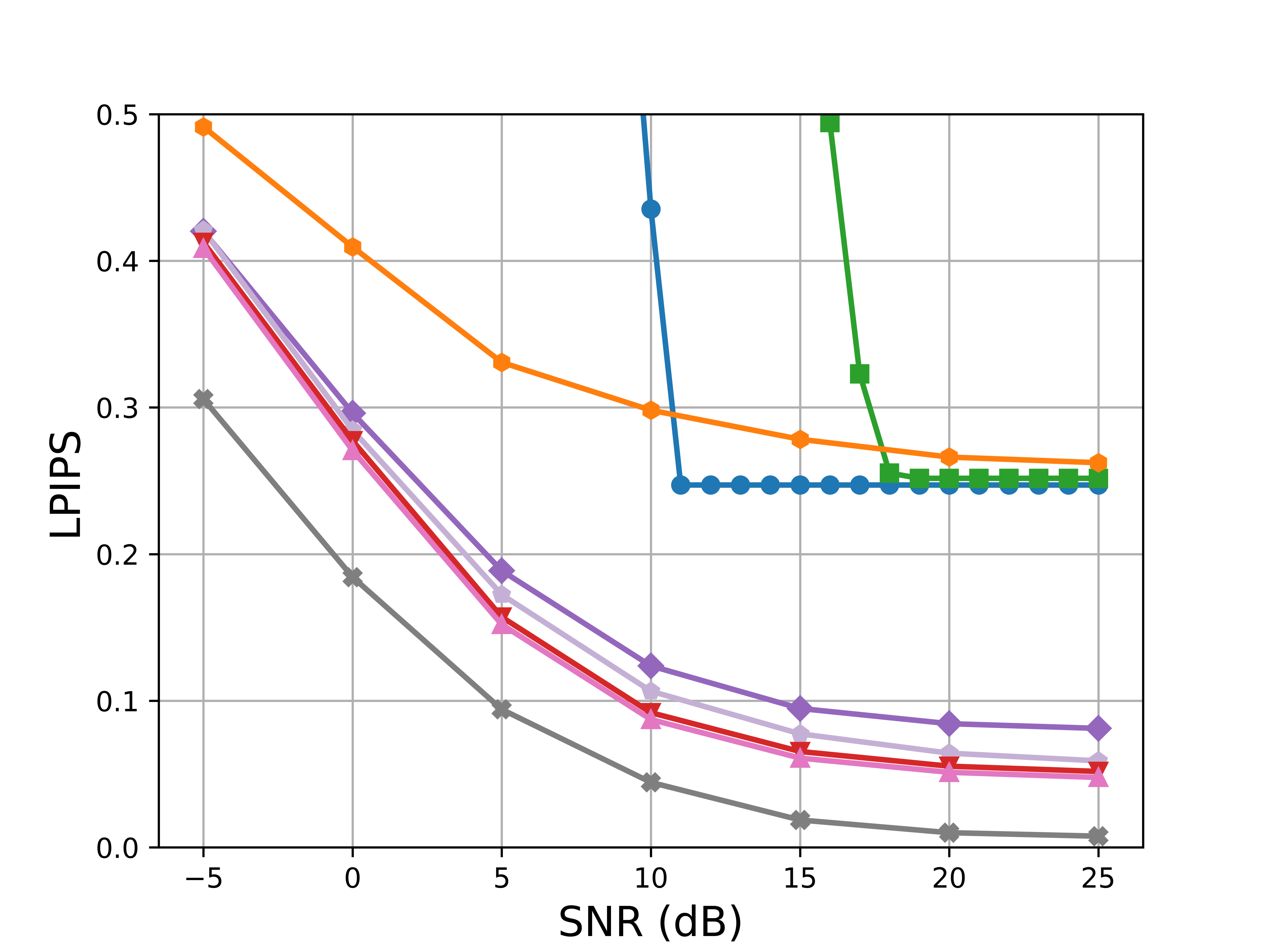} \\
        {\footnotesize (a) AWGN Channel}
        \label{fig8a}
    \end{minipage}


    \begin{minipage}{\textwidth}
        \centering
        \includegraphics[width=0.27\textwidth]{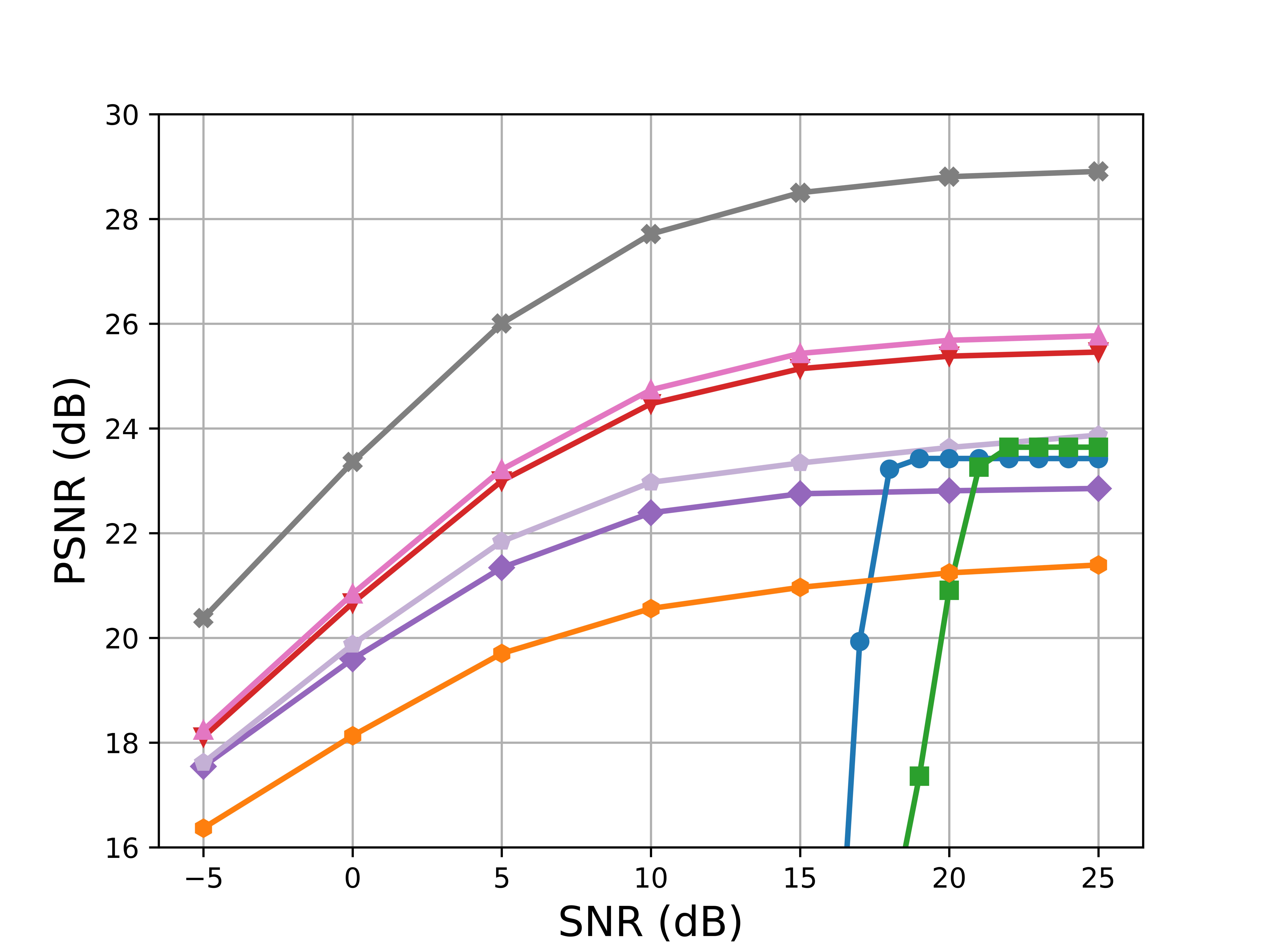} \hspace{10pt}
        \includegraphics[width=0.27\textwidth]{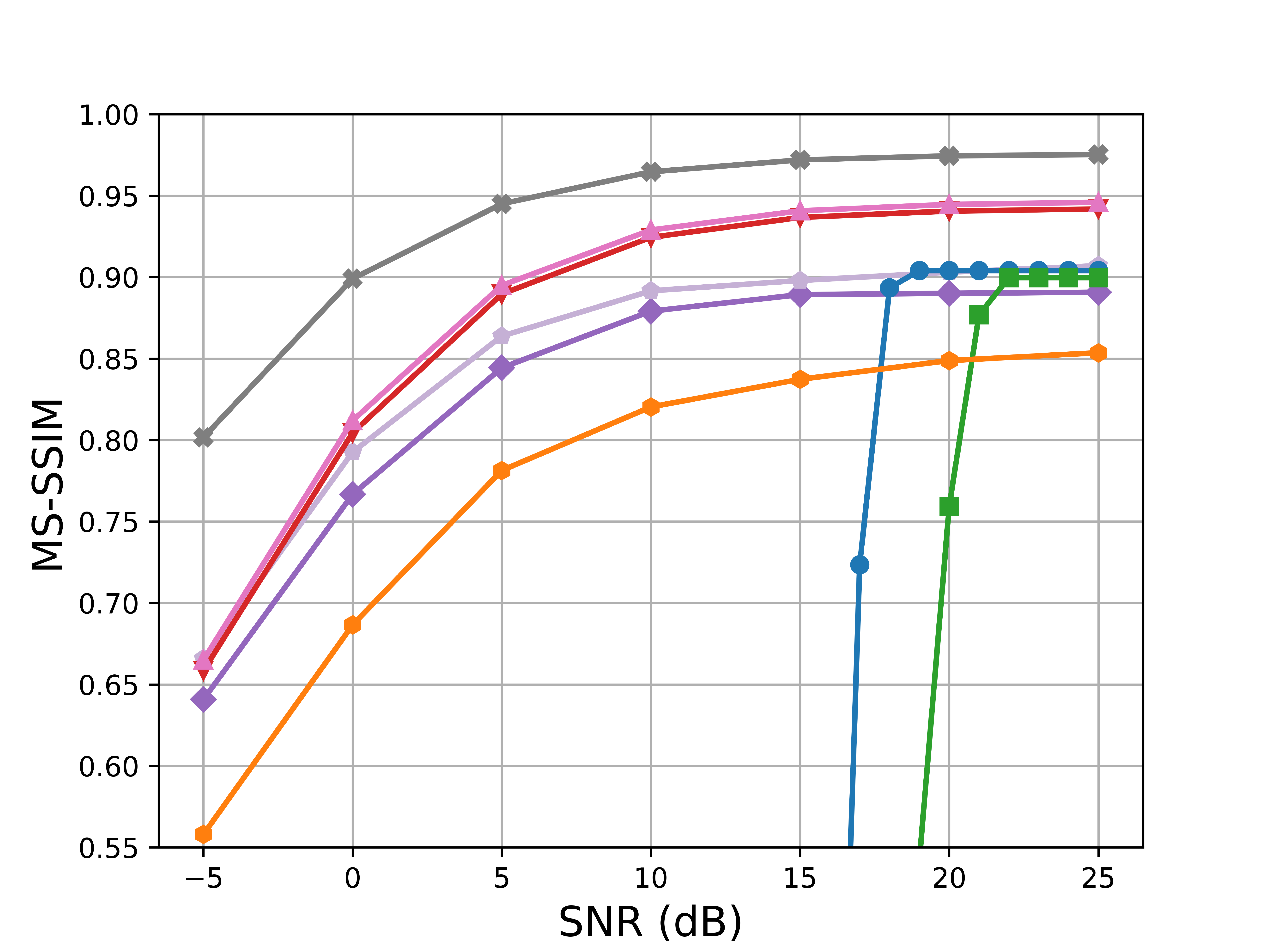} \hspace{10pt}
        \includegraphics[width=0.27\textwidth]{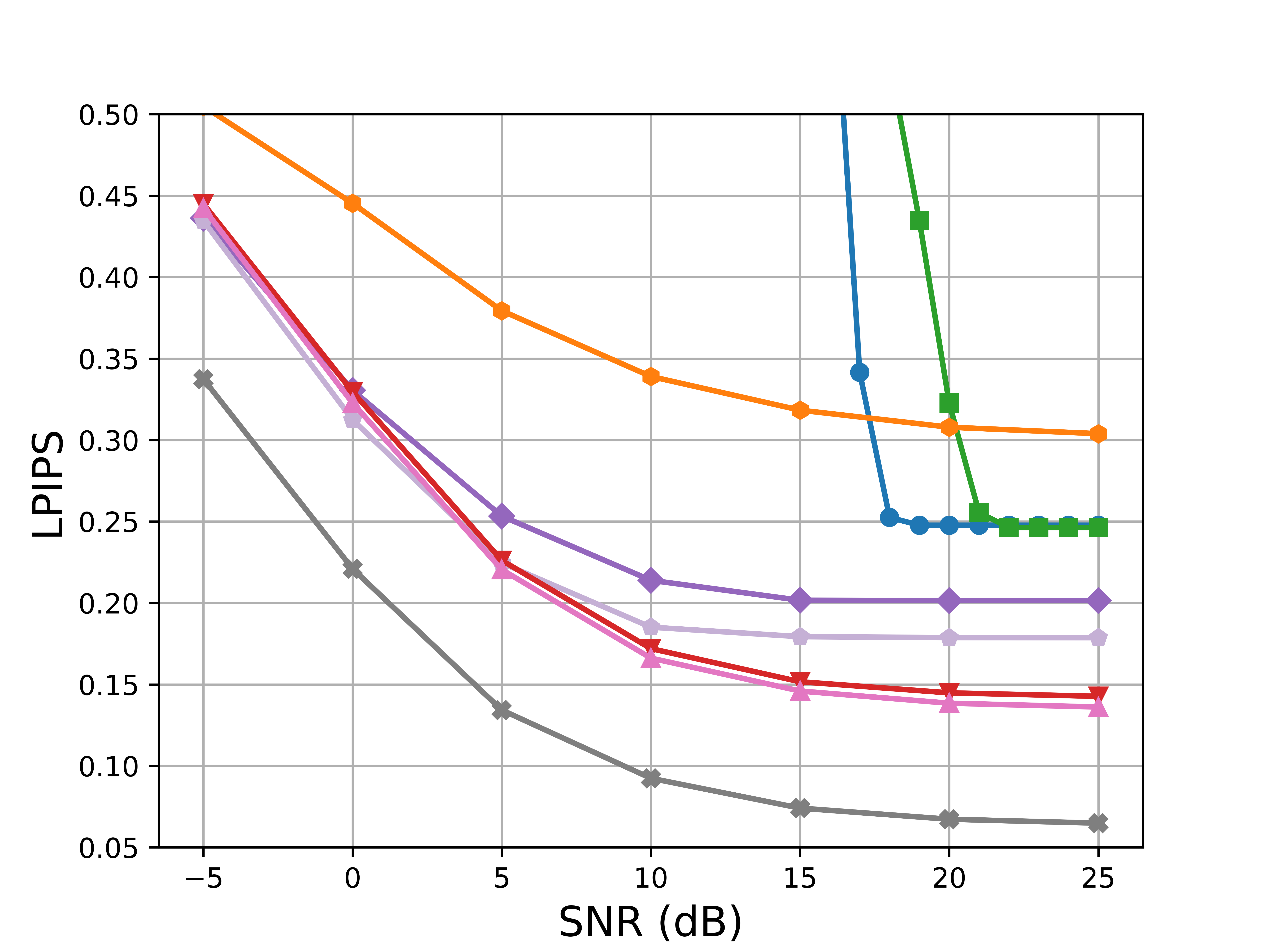} \\
        {\footnotesize (b) Rayleigh Channel}
        \label{fig8b}
    \end{minipage}


    \begin{minipage}{\textwidth}
        \centering
        \includegraphics[width=0.27\textwidth]{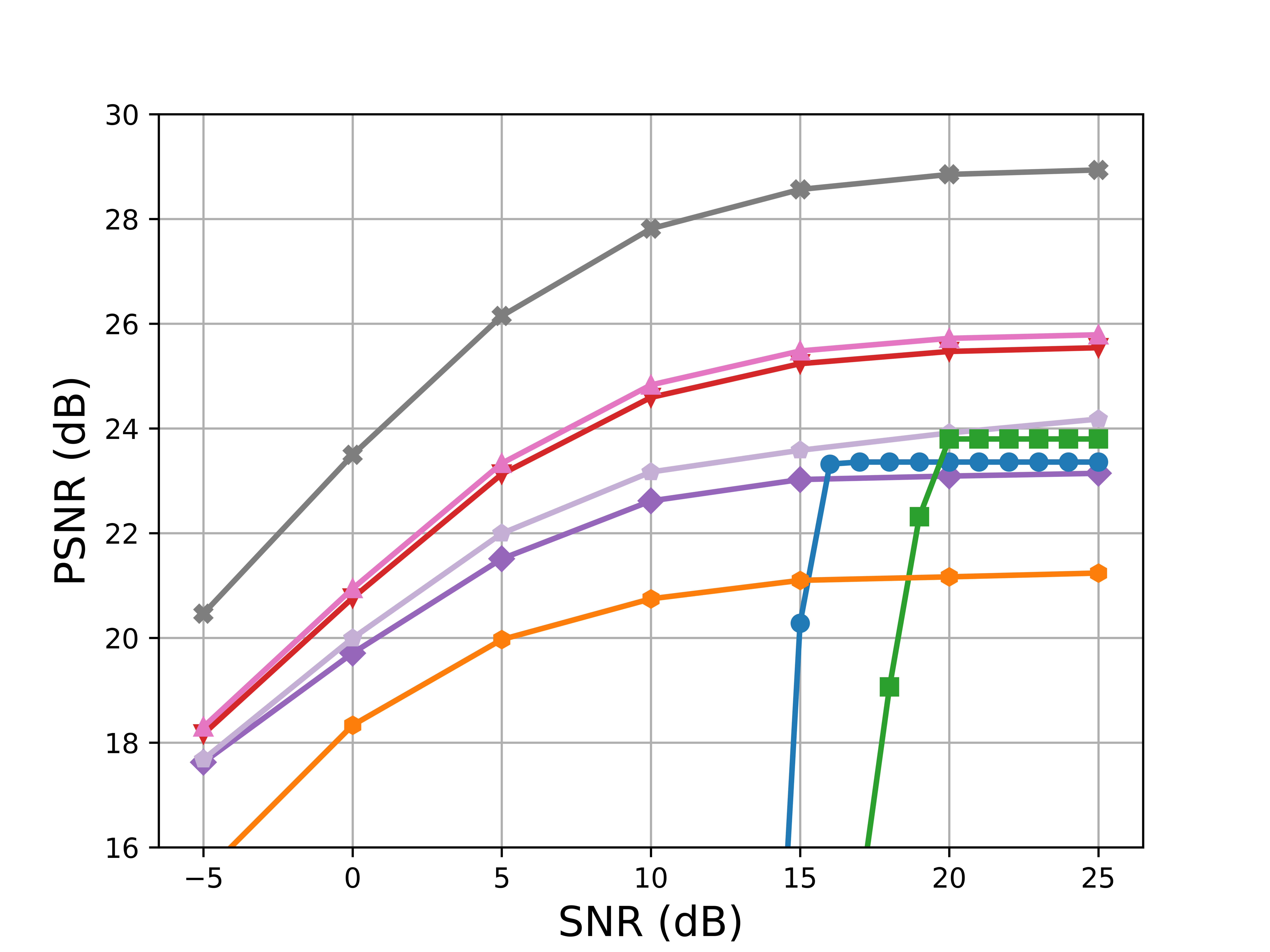} \hspace{10pt}
        \includegraphics[width=0.27\textwidth]{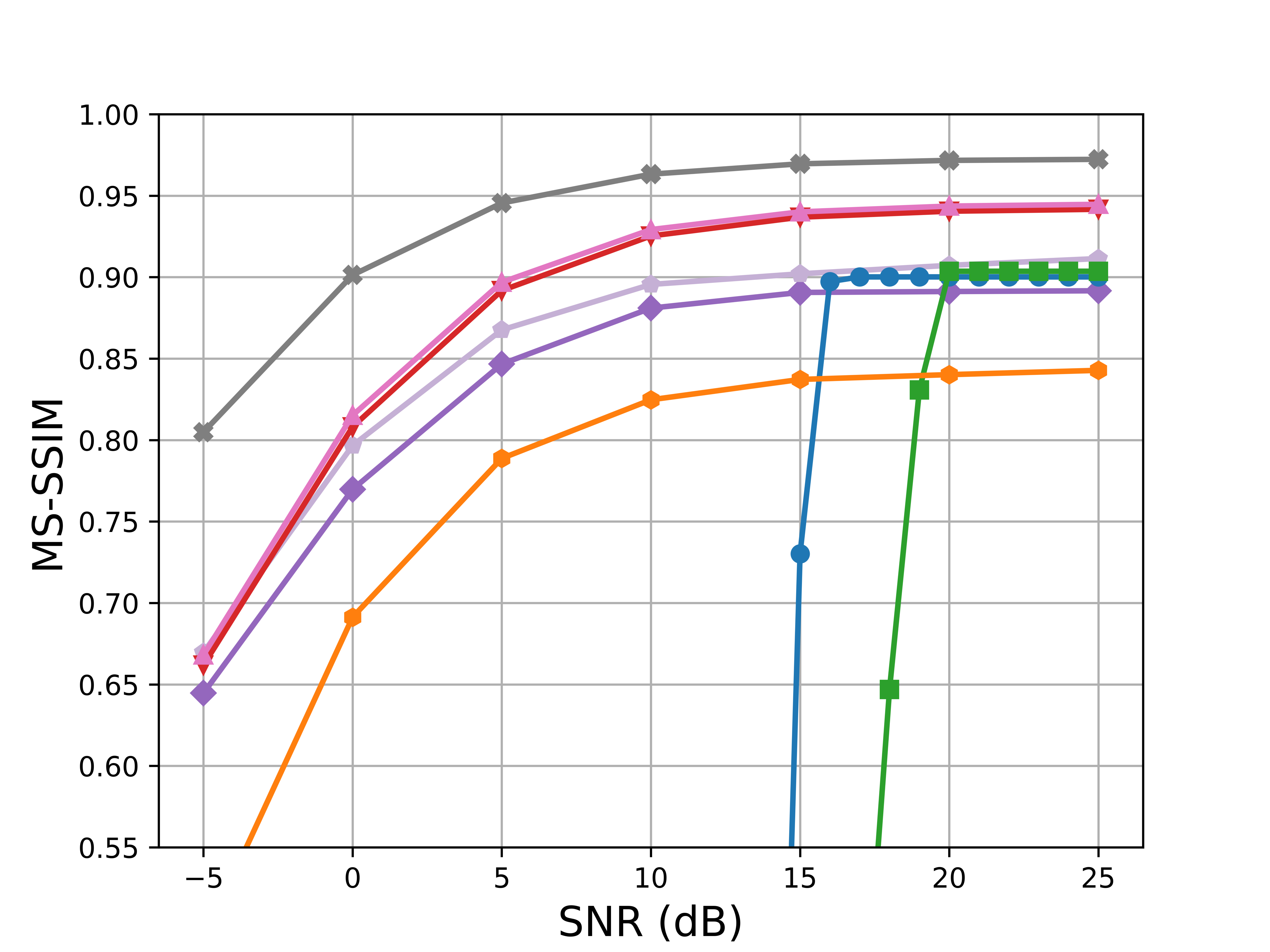} \hspace{10pt}
        \includegraphics[width=0.27\textwidth]{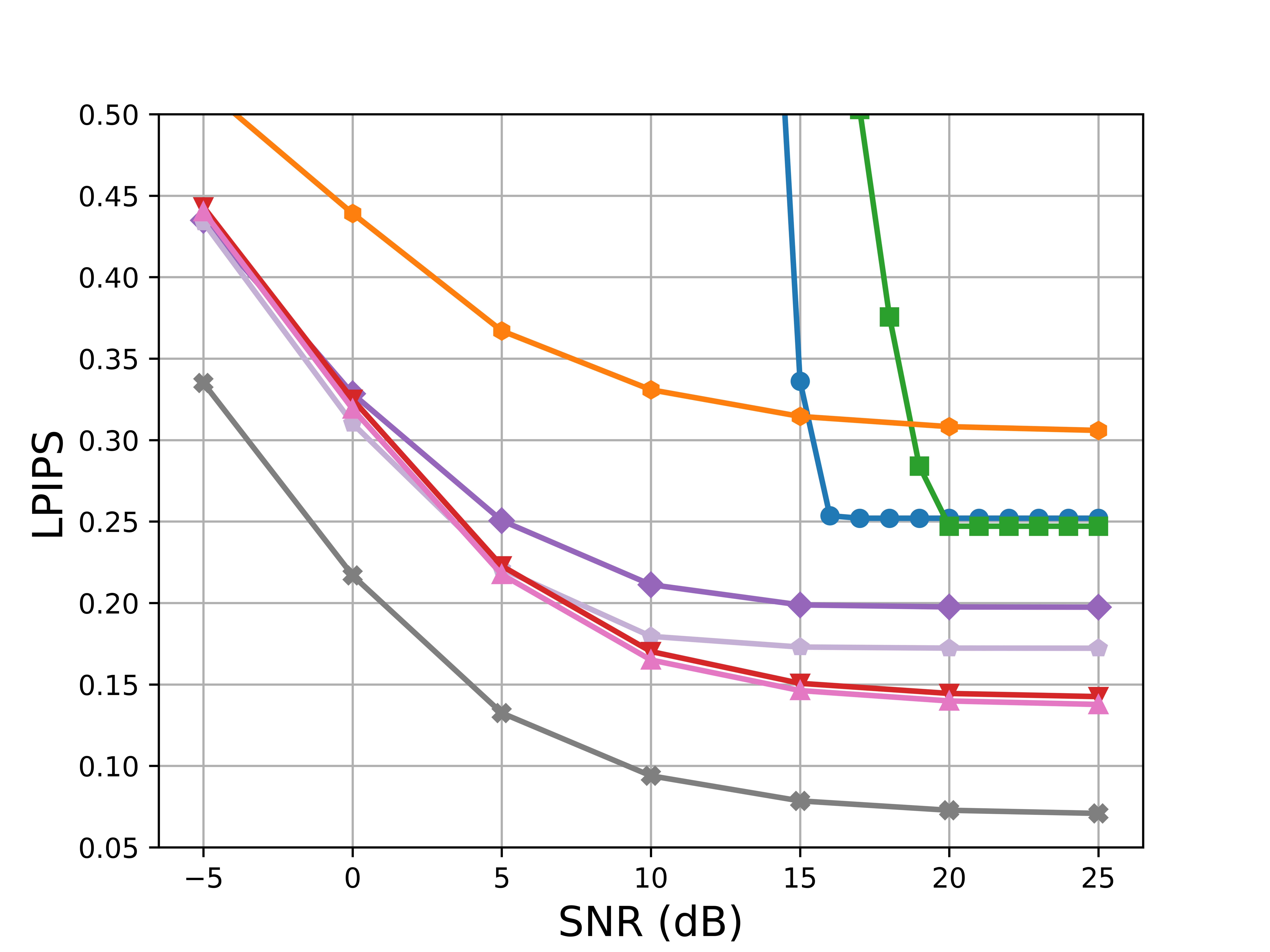} \\
        {\footnotesize (c) Rician Channel}
        \label{fig8c}
    \end{minipage}
    
    \vspace{2pt}
    
    \begin{minipage}{0.5\textwidth}
        \centering
        \includegraphics[width=\linewidth]{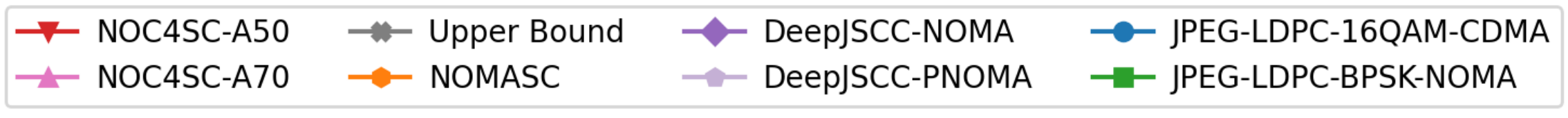}
    \end{minipage}

    \caption{Performance comparison of different schemes on the CIFAR10 dataset in the three-user communication scenario under (a) AWGN, (b) Rayleigh, and (c) Rician channels.}
    \label{fig8}
    
\end{figure*}

\captionsetup[table]{justification=centering, labelsep=space, textfont=sc} 
\begin{table}[t]
    \renewcommand\arraystretch{1.6}
    \setlength{\tabcolsep}{10pt}
    \caption{ \\ Three-User Semantic Feature Decoding Using Different Non-Orthogonal Codewords\label{tab2}}
    \centering
    \begin{tabular}{cccc}
        \hline
        \hline
         & $g(\bm{z_1}, \bm{c_1})$ & $g(\bm{z_1}, \bm{c_2})$ & $g(\bm{z_1}, \bm{c_3})$\\
        \hline
        PSNR/dB $\uparrow$ & $\bm{28.488}$ & 5.399 & 6.026 \\
        MS-SSIM $\uparrow$ & $\bm{0.984}$  & 0.137 & 0.117 \\
        LPIPS $\downarrow$ & $\bm{0.080}$  & 0.620 & 0.624 \\
        \hline
        \hline
        & $g(\bm{z_2}, \bm{c_1})$ & $g(\bm{z_2}, \bm{c_2})$ & $g(\bm{z_2}, \bm{c_3})$\\
        \hline
        PSNR/dB $\uparrow$ & 5.348 & $\bm{29.418}$ & 5.800 \\
        MS-SSIM $\uparrow$ & 0.137  & $\bm{0.984}$ & 0.112 \\
        LPIPS $\downarrow$ & 0.614  & $\bm{0.060}$ & 0.625 \\
        \hline
        \hline
        & $g(\bm{z_3}, \bm{c_1})$ & $g(\bm{z_3}, \bm{c_2})$ & $g(\bm{z_3}, \bm{c_3})$\\
        \hline
        PSNR/dB $\uparrow$ & 5.393 & 7.895 & $\bm{30.456}$ \\
        MS-SSIM $\uparrow$ & 0.137 & 0.127 & $\bm{0.986}$ \\
        LPIPS $\downarrow$ & 0.614 & 0.697 & $\bm{0.056}$ \\
        \hline
        \hline
    \end{tabular}
\end{table}

\textit{4) Simulation Hyperparameters:} For the low-resolution CIFAR-10 dataset, two stages are utilized, with the corresponding numbers of ST blocks configured as [2, 4], the window size of 2, and the NOC length of 128. For the AFHQ dataset, four stages are utilized, with the corresponding numbers of ST blocks configured as [2, 2, 6, 2], the window size of 8, and the NOC length of 64. All implementations were executed in Python 3.8.12 and PyTorch 1.11, with CUDA 11.6, across two NVIDIA V100 GPUs. Some other perparameters in the simulatin are listed in Table \ref{tab5}.

\subsection{Results Analysis}

\textit{1) Distinguishable Semantic Feature Domains:} To demonstrate the separability of semantic features, we utilized the t-Distributed Stochastic Neighbor Embedding (t-SNE) \cite{ref53} technique to project the high-dimensional semantic features into a two-dimensional space, which is particularly effective for visualizing complex data distributions while preserving the local structure of the original high-dimensional feature space. Fig.\ref{fig5a} illustrates the two-dimensional t-SNE projection of high-dimensional semantic features $\bm{z_1}$, $\bm{z_2}$, and $\bm{z_3}$ extracted from the NOC4SC system under the three-user scenario on the CIFAR-10 dataset. The visualization clearly demonstrates the separability of the semantic feature subspaces corresponding to different TXs. Specifically, the feature clusters associated with each user are well-separated in the projected space, indicating that the NOC-based multi-user access mechanism effectively maps the semantic features of different users into distinguishable subspaces, thereby mitigating inter-user interference. 

\begin{figure*}[t]
    \centering
    \begin{minipage}{\textwidth}
        \centering
        \includegraphics[width=0.27\textwidth]{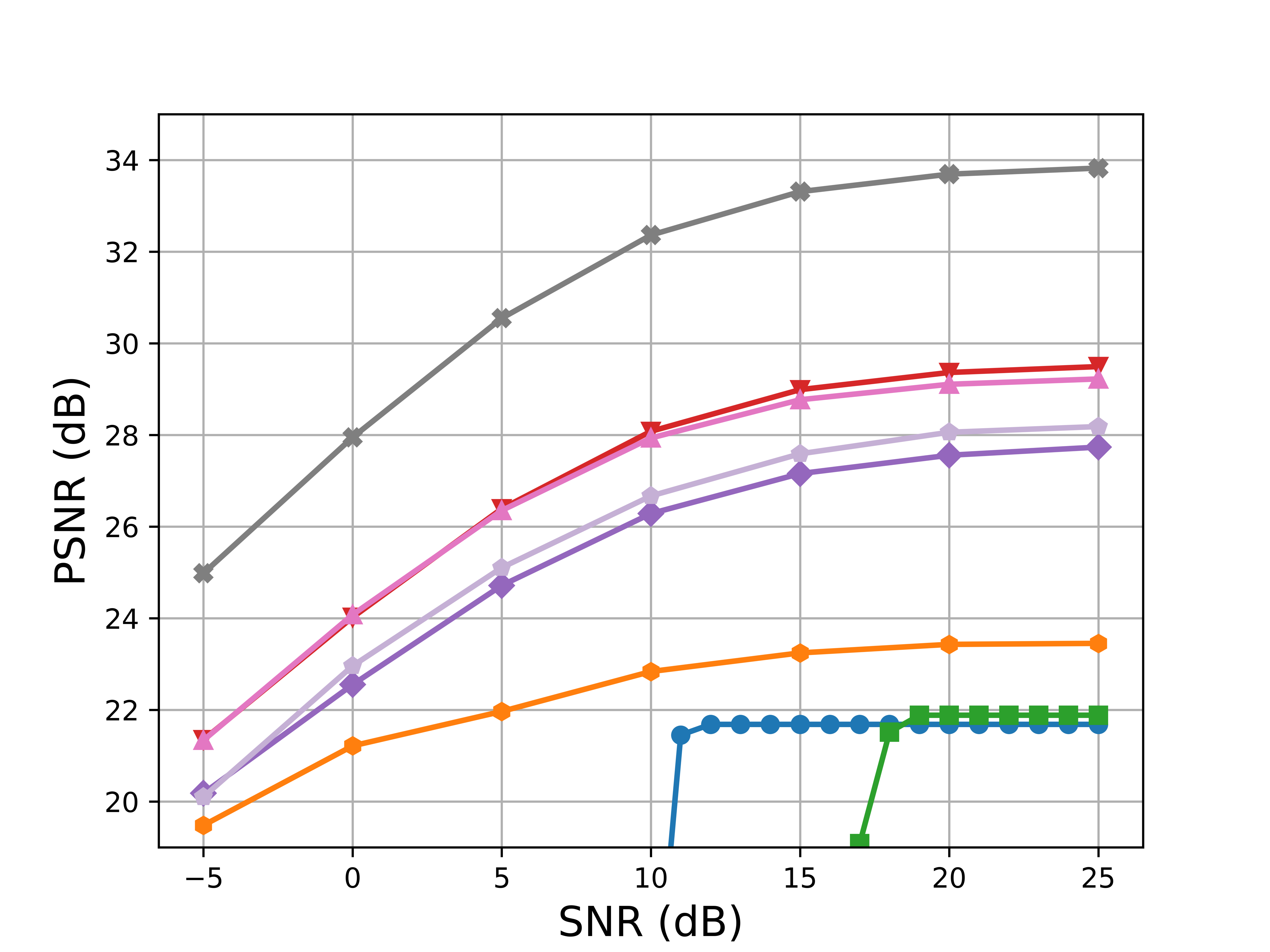} \hspace{10pt}
        \includegraphics[width=0.27\textwidth]{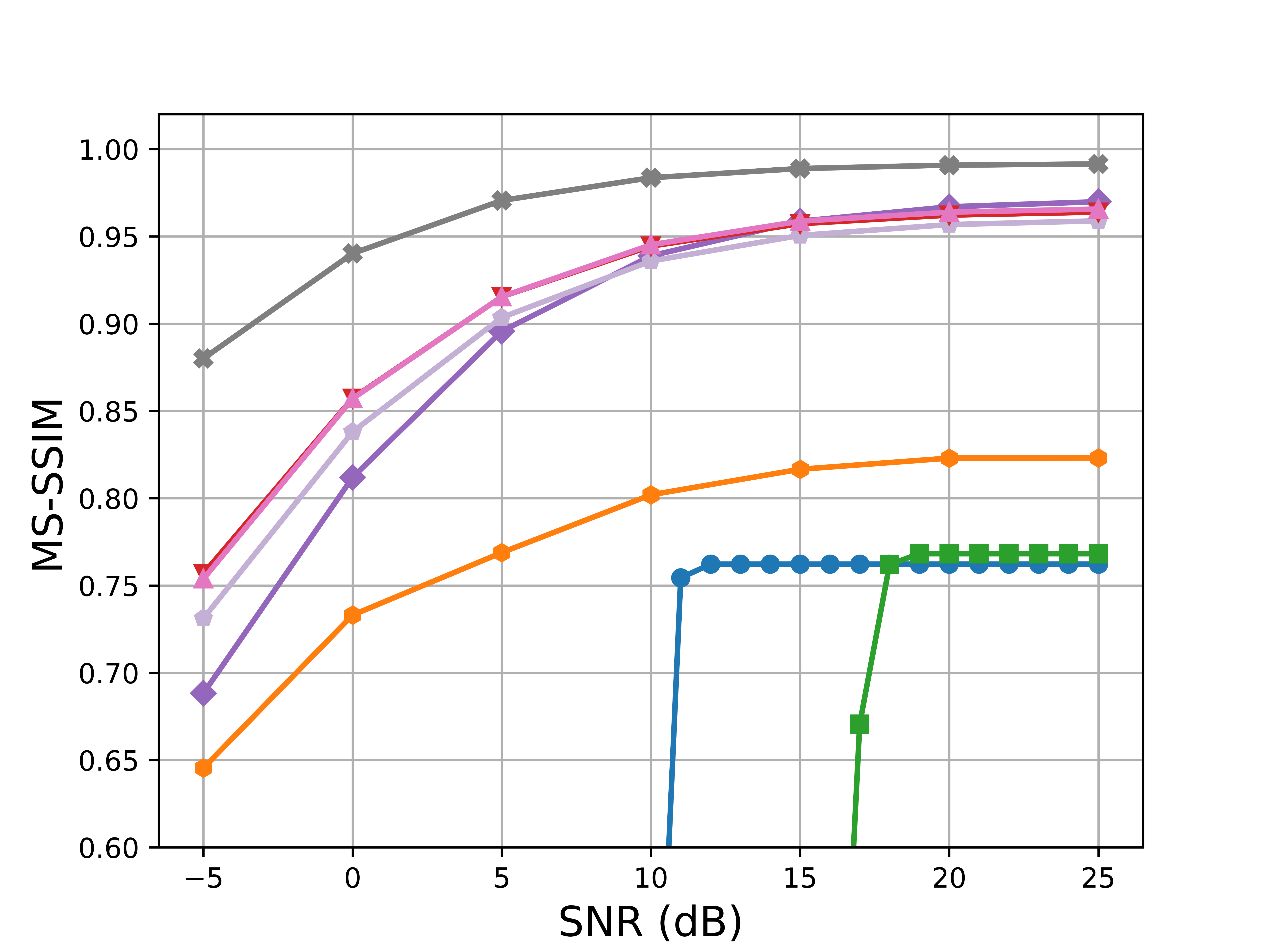} \hspace{10pt}
        \includegraphics[width=0.27\textwidth]{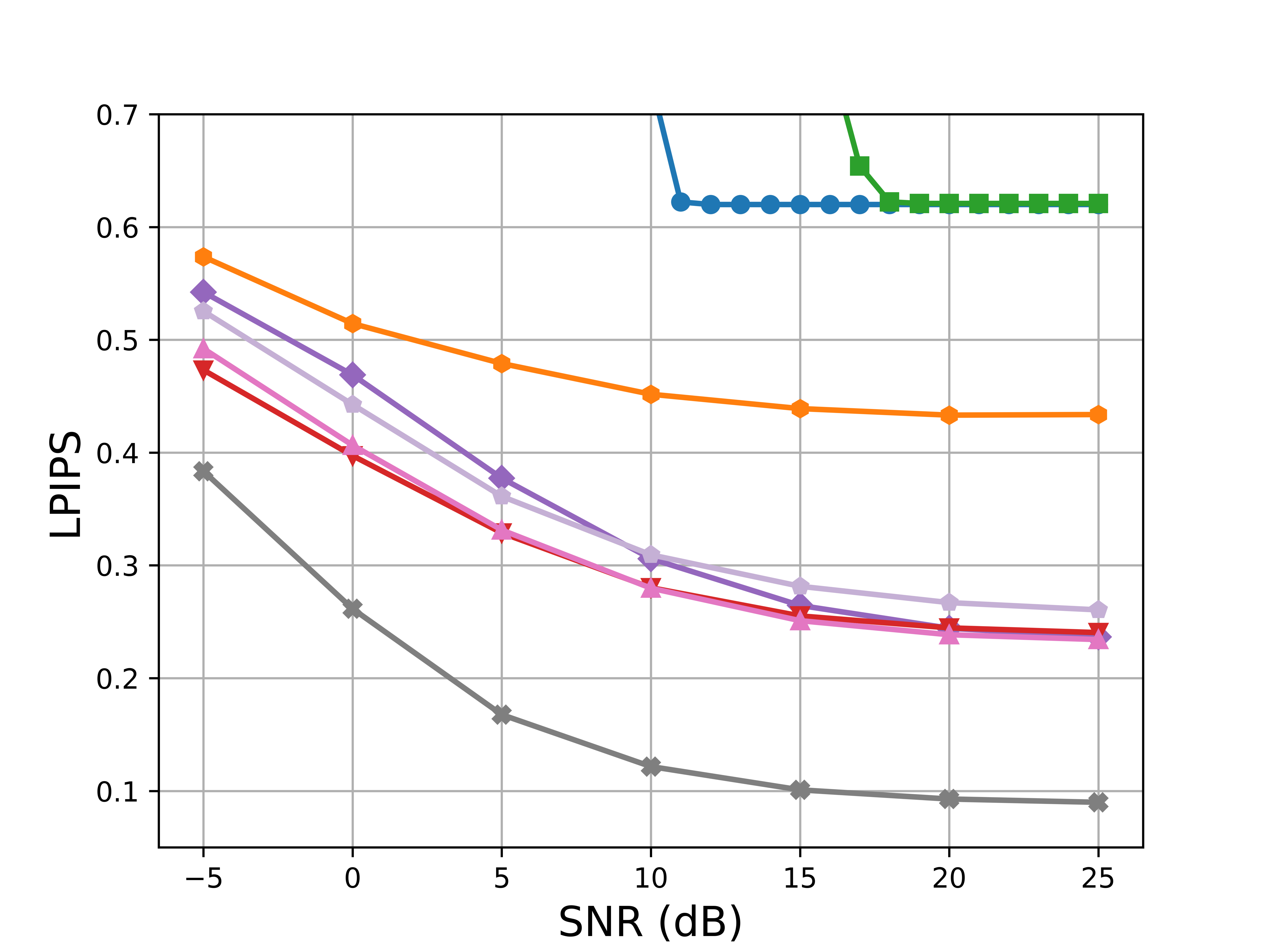} \\
        {\footnotesize (a) AWGN Channel}
        \label{fig9a}
    \end{minipage}

    \begin{minipage}{\textwidth}
        \centering
        \includegraphics[width=0.27\textwidth]{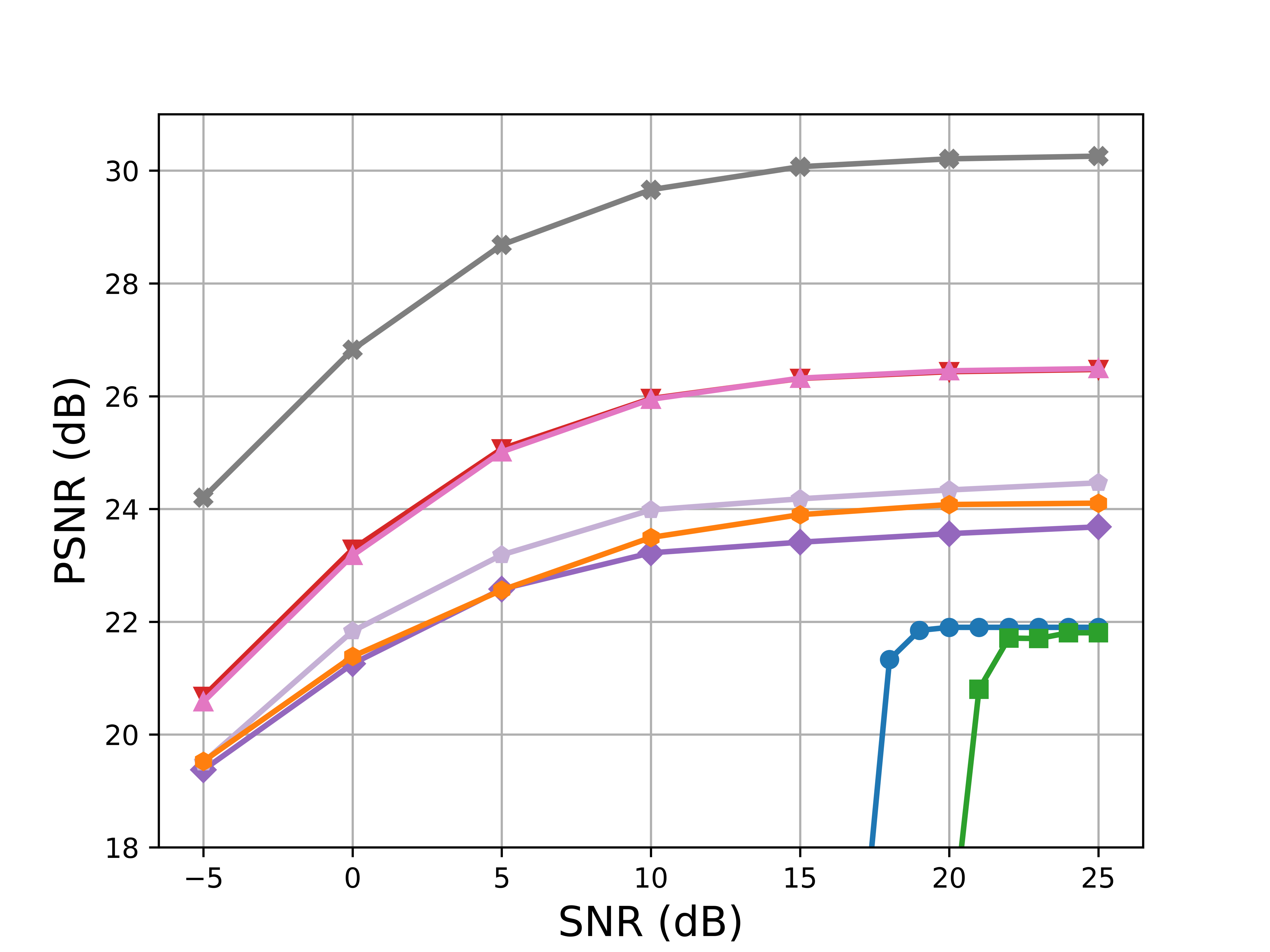} \hspace{10pt}
        \includegraphics[width=0.27\textwidth]{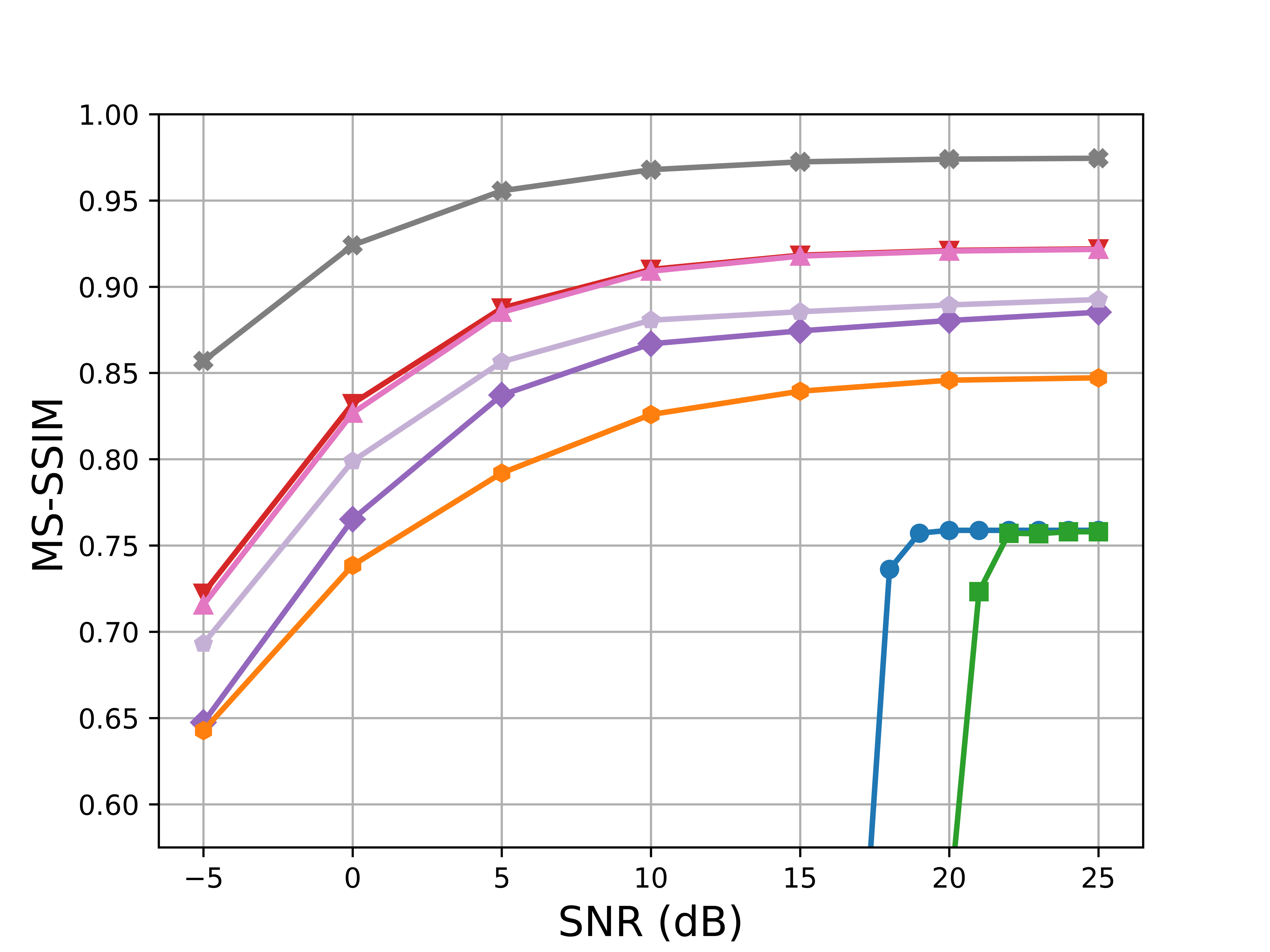} \hspace{10pt}
        \includegraphics[width=0.27\textwidth]{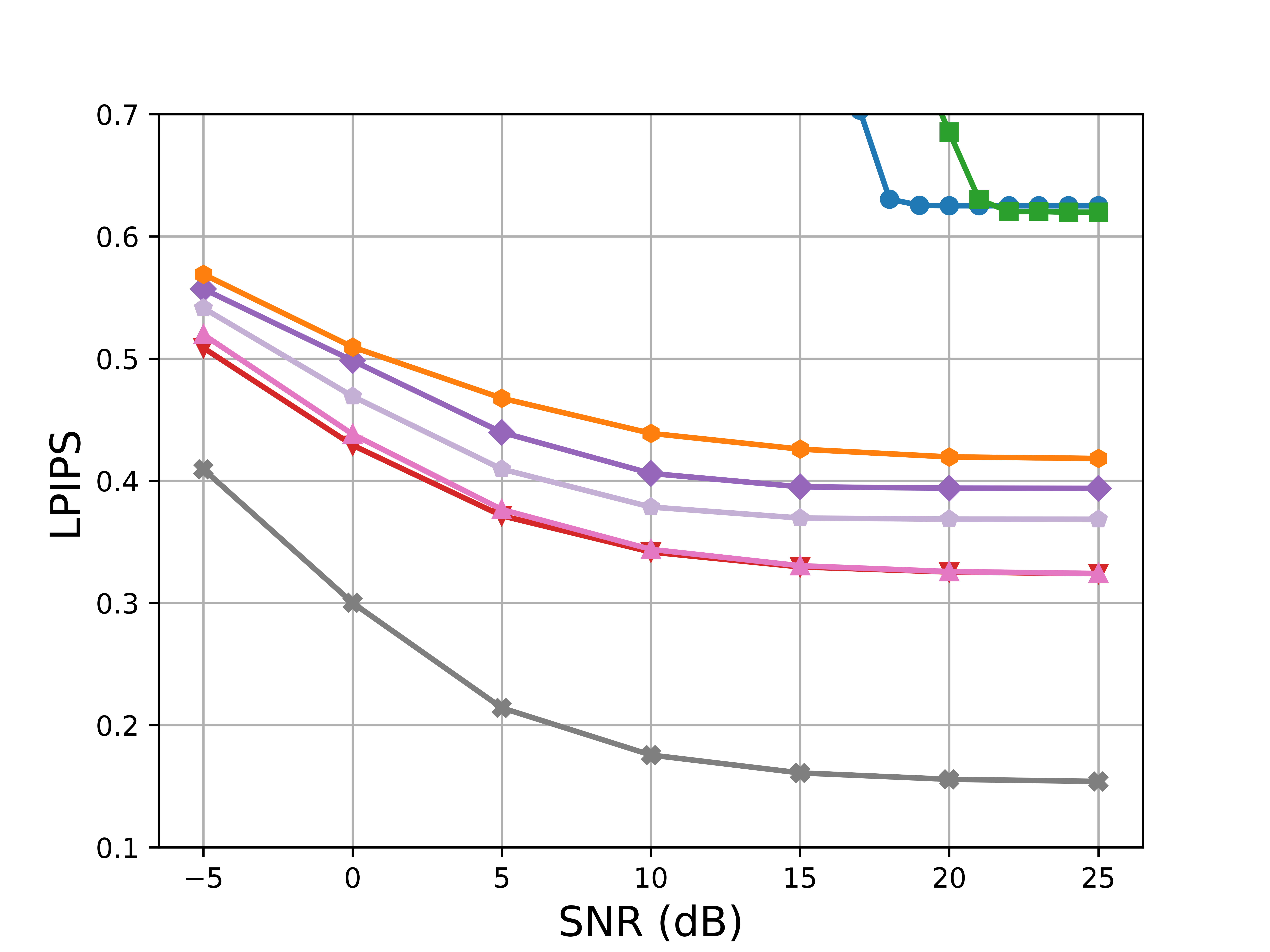} \\
        {\footnotesize (b) Rayleigh Channel}
        \label{fig9b}
    \end{minipage}

    \begin{minipage}{\textwidth}
        \centering
        \includegraphics[width=0.27\textwidth]{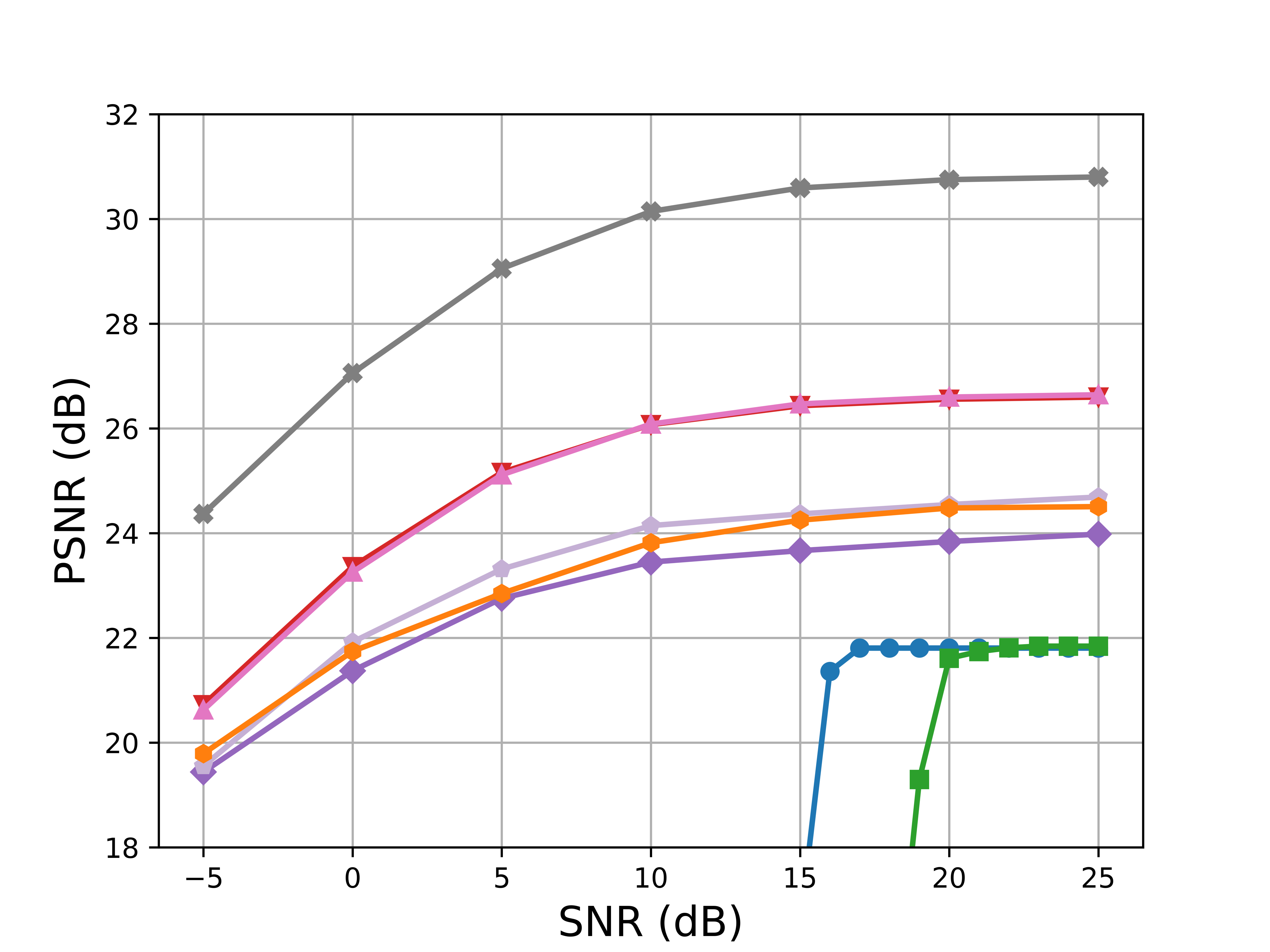} \hspace{10pt}
        \includegraphics[width=0.27\textwidth]{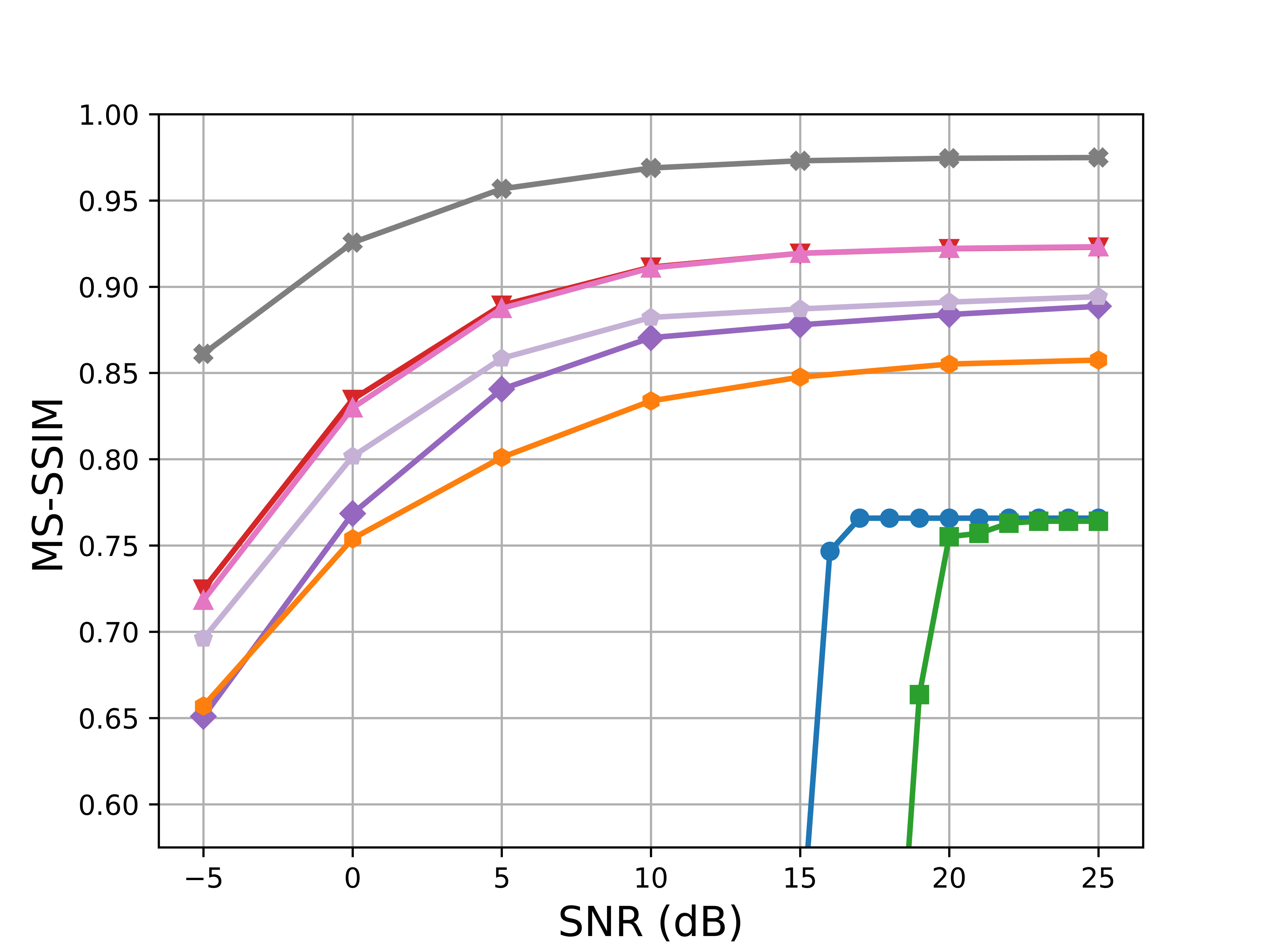} \hspace{10pt}
        \includegraphics[width=0.27\textwidth]{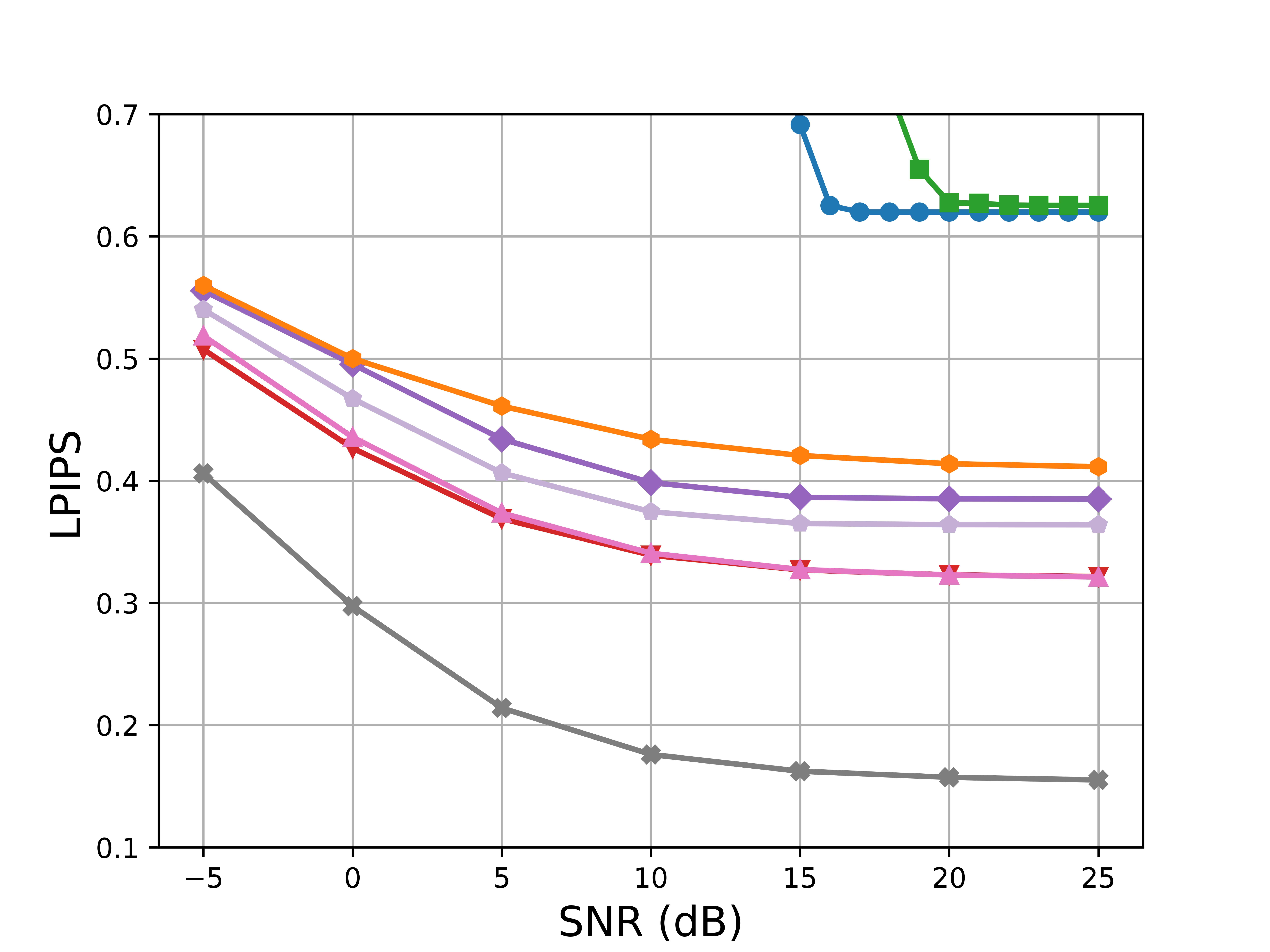} \\
        {\footnotesize (c) Rician Channel}
        \label{fig9c}
    \end{minipage}
    
    \vspace{2pt}
    
    \begin{minipage}{0.5\textwidth}
        \centering
        \includegraphics[width=\linewidth]{./figure/Legend.pdf}
    \end{minipage}

    \caption{Performance comparison of different schemes on the AFHQ dataset in the three-user communication scenario under (a) AWGN, (b) Rayleigh, and (c) Rician channels.}
    \label{fig9}
    
\end{figure*}

\textit{2) Independence of Semantic Features Across Users:} To futher investigate the orthogonality properties of the semantic feature, table \ref{tab1} presents the normalized inner products and the corresponding angles between different users' semantic feature vectors $\bm{z_i}$ on the CIFAR-10 dataset. The inner products between normalized features are close to zero, and the angles between the features are all approximately $90^{\circ}$, which strongly evidences that the semantic subspaces assigned to different users are effectively separated, thus supporting interference suppression.

Table \ref{tab2} presents the decoding performance of semantic features $\bm{z}$ in the three-user communication scenario using different NOCs. Here, $g(\bm{z_i}, \bm{C_j})$ denotes the decoding of the semantic feature $\bm{z_i}$ of the $i$-th user using the codeword $\bm{C_j}$ assigned to the $j$-th user. When the correct codeword $\bm{C_i}$ is used, i.e., $g(\bm{z_i}, \bm{C_i})$, the reconstructed quality remains high, with PSNR consistently above 31 dB. In contrast, applying mismatched codewords results in a dramatic performance drop, with PSNR falling to around 5 dB and MS-SSIM below 0.137. These results confirm not only the separability of user-specific semantic features enabled by the proposed NOC design, but also imply a security-enhancing property: only the correct codeword enables successful decoding, inherently preventing unauthorized access to other users's semantic feature.

\textit{3) Analysis of Results for Image Reconstruction:} Fig.\ref{fig8} and Fig.\ref{fig9} compare the performance of different communication schemes on the CIFAR-10 and AFHQ datasets, respectively, under a three-user semantic communication scenario. The evaluations are conducted over AWGN, Rayleigh, and Rician channels to jointly assess pixel-level fidelity and perceptual quality. Across all SNR ranges and channel conditions, the proposed NOC4SC framework consistently outperforms existing baselines, including DeepJSCC-NOMA, DeepJSCC-PNOMA, and NOMASC. The superiority of NOC4SC is particularly evident in the LPIPS metric, where lower values denote better perceptual alignment, and in the PSNR and MS-SSIM metrics, which reflect enhanced reconstruction accuracy and structural consistency. 

Specifically, under the AWGN channel (Fig.~\ref{fig8}(a)), both NOC4SC-A50 and NOC4SC-A70 configurations achieve notable gains, exceeding the performance of DeepJSCC-PNOMA by up to 1.22 dB in PSNR at high SNR. Even under Rayleigh fading conditions (Fig.~\ref{fig8}(b)), where multipath fading and amplitude fluctuation degrade the transmission quality, NOC4SC maintains a clear performance advantage with a PSNR gain of 2.14 dB over DeepJSCC-PNOMA. The results under Rician channels (Fig.~\ref{fig8}(c)) further confirm that NOC4SC remains robust in mixed line-of-sight and multipath environments, demonstrating consistent superiority across all metrics. A similar performance trend is observed on the AFHQ dataset (Fig.~\ref{fig9}), demonstrating that the proposed NOC4SC framework generalizes well across different data distributions and semantic domains. 

In contrast, traditional coding and modulation approaches, such as NOMA and CDMA-based schemes, though competitive at high SNR, exhibit the characteristic “cliff effect” where reconstruction quality deteriorates sharply at low SNR due to their limited robustness to noise and fading. Overall, these results demonstrate that NOC4SC not only ensures high reconstruction quality but also exhibits strong adaptability and resilience across diverse wireless channel conditions, thereby validating its effectiveness for practical multi-user semantic communication.

\begin{figure*}[t]
\centering
\begin{minipage}{0.325\textwidth}
        \centering
        \subfloat[(a) PSNR)]{\includegraphics[width=\linewidth]{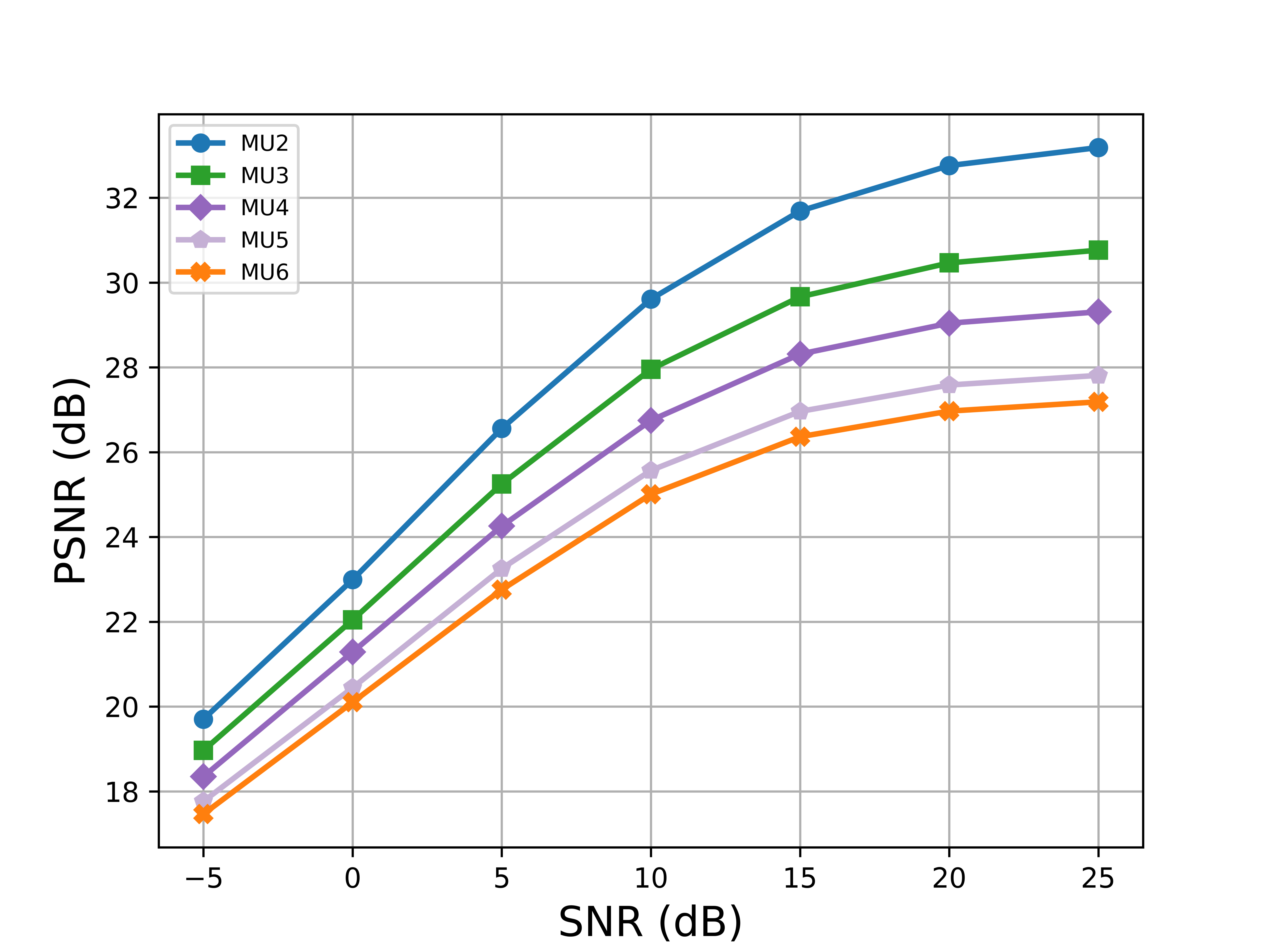}%
        \label{fig3a}}
\end{minipage}
\begin{minipage}{0.325\textwidth}
        \centering
        \subfloat[(b) MS-SSIM]{\includegraphics[width=\linewidth]{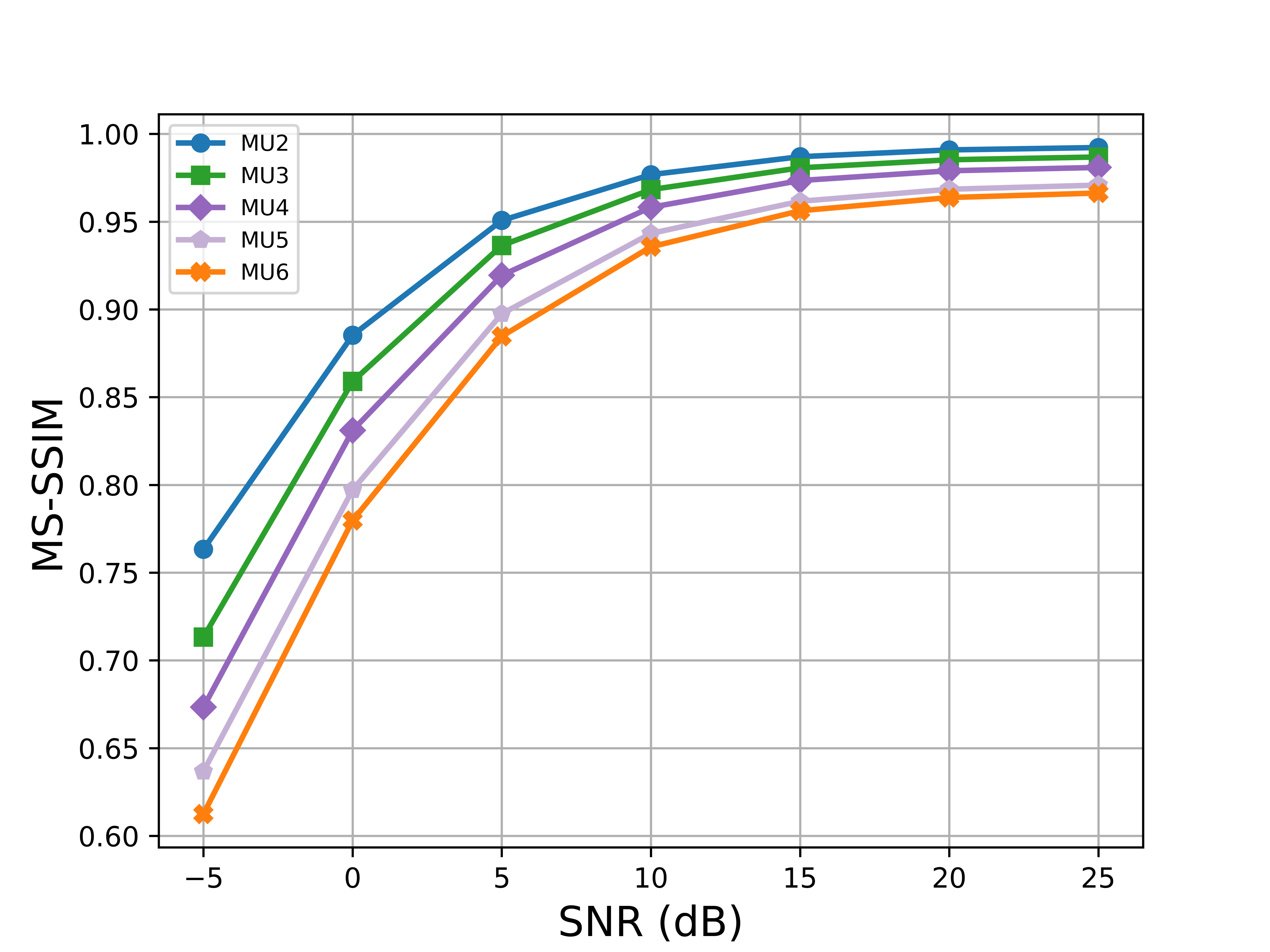}%
        \label{fig3b}}
\end{minipage}
\begin{minipage}{0.325\textwidth}
        \centering
        \subfloat[(c) LPIPS]{\includegraphics[width=\linewidth]{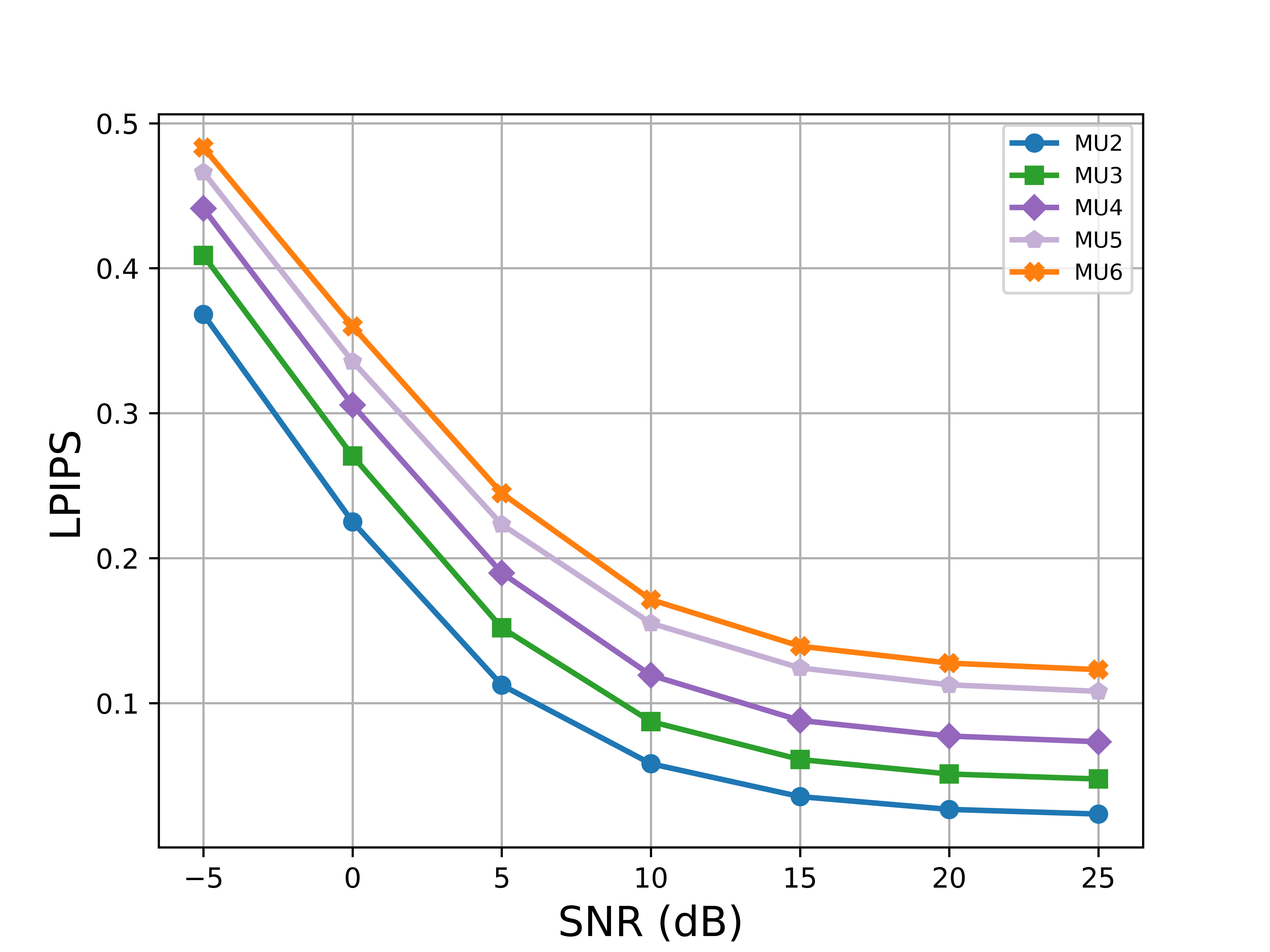}%
        \label{fig3c}}
\end{minipage}
\caption{Performance comparison of semantic communication systems with varying numbers of users on the CIFAR-10 dataset under AWGN channels. Noting: MU $\bm{N}$ denotes a multi-user scenario involving $\bm{N}$ users.}
\label{fig10}
\end{figure*}

\captionsetup[table]{justification=centering, labelsep=space, textfont=sc} 
\begin{table*}[t]
    \renewcommand\arraystretch{1.6}
    \setlength{\tabcolsep}{6pt}
    \caption{ \\ Computational Complexity and Latency Comparison on the AFHQ dataset.\label{tab8}}
    \centering
    \begin{tabular}{ccccccc}
        \hline
        \hline
        \multicolumn{2}{c}{Transmission scheme} & Parameters (M) & FLOPs & Encoding latency & Decoding latency & Inference time\\
        \hline
        \multicolumn{2}{c}{JPEG-LDPC-BPSK-NOMA} & \textbf{-} & \textbf{-} & $>$160 ms & $>$22.11 s & $>$22.14 s\\
        \multicolumn{2}{c}{JPEG-LDPC-16QAM-CDMA} & \textbf{-} & \textbf{-} & $>$260 ms & $>$14.99 s & $>$15.29 s\\
        \multicolumn{2}{c}{NOMASC} & 28.21 & 147.86 G & 30.50 ms & 42.11 ms & 81.93 ms\\
        \multicolumn{2}{c}{\textbf{NOC4SC}} & \textbf{42.53} & \textbf{148.60 G} & \textbf{19.37 ms} & \textbf{26.37 ms} & \textbf{78.87 ms}\\
        \hline
        \hline
    \end{tabular}
\end{table*}

\textit{4) Scalability Analysis:} To evaluate the scalability of the proposed NOC4SC framework, we conducted simulations on the CIFAR-10 dataset under AWGN channels, as illustrated in Fig.\ref{fig10}. The experiments considered multi-user scenarios ranging from two to six users. The reconstruction performance—measured by PSNR, MS-SSIM, and LPIPS—exhibits a gradual decline with an increasing number of users. This degradation stems from the fact that, although the theoretical design of NOC4SC enforces mutual orthogonality among user-specific feature subspaces, perfect orthogonality cannot be achieved in practice due to finite-dimensional feature representations and optimization limitations of neural networks. Consequently, as the number of users grows, the inter-user interference within the shared semantic feature space becomes more pronounced, leading to a reduction in perceptual quality. Nevertheless, the proposed NOC4SC scheme preserves satisfactory reconstruction performance, confirming its ability to support efficient multi-user semantic transmission without bandwidth expansion.

\textit{5) Complexity Analysis:} We evaluate the computational complexity and processing latency of different transmission schemes under 10 SNR on the Rayleigh fading channel on the AFHQ dataset show the mertics in Table \ref{tab8}, including model parameters, FLOPs, latency ,and inference time. We conducted 100 images on the AFHQ dataset with a batch size of 1 to obtain the average encoding, decoding and inference time per image. It can be observed that the proposed NOC4SC scheme achieves significantly faster runtime compared to classical methods such as NOMA and CDMA, primarily due to the elimination of LDPC coding and the computational overhead associated with SIC operations in NOMA. NOC4SC exhibits lower computational latency than the state-of-the-art NOMASC scheme by sharing network parameters across users and enabling parallel semantic reconstruction.

\section{Conclusion}
In this paper, we proposed NOC4SC, a novel multi-user access paradigm designed to address the challenges of bandwidth limitations and inter-user interference. NOC4SC provides a scalable approach to enhancing spectral efficiency in multi-user SemCom, which mitigates interference and ensures data security through approximately orthogonal semantic feature extraction. In the NOC4SC system, we introduced the NSM block, which enables users to independently extract semantic features while sharing network parameters. Extensive experimental results demonstrate the proposed NOC4SC system achieves comparable performance versus the baseline methods, particularly with low-to-medium SNRs.

While the NOC4SC system has achieved promising performance across image datasets, the current study is limited to image transmission scenarios. Future research will extend the framework to support multi-user SemCom diverse data modalities such as speech and video. In addition, we will further investigate both asymmetric and symmetric quantization and rigorously characterize their associated complexity–accuracy.

\end{document}